\documentclass[]{article}
\usepackage{jheppub}
\usepackage{hyperref}
\usepackage[most]{tcolorbox}
\usepackage{mdframed}
\usepackage{lipsum}
\usepackage{varwidth}

\tcbset{
  highlight math/.style={notitle,nophantom,colframe=red,colback=yellow!25!white}
}

\usepackage{setspace}
\usepackage{physics}
\usepackage{dsfont}
\usepackage{tensor}
\usepackage[normalem]{ulem}
\usepackage{xcolor}
\usepackage{graphicx}
\usepackage[export]{adjustbox}
\usepackage{mathtools}
\usepackage{mathbbol}
\usepackage[utf8]{inputenc}
\usepackage{amsmath}
\usepackage{bbold}
\usepackage{breqn}
\usepackage{blindtext}
\usepackage{float}
\usepackage[english]{babel}
\usepackage{amssymb,amsthm}
\usepackage{ragged2e}
\usepackage{etoolbox}
\usepackage{lipsum}
\usepackage{multirow}
\usepackage{tcolorbox}
\usepackage{latexsym}
\usepackage{enumitem}

\usepackage{titlesec}
\usepackage{url}
\usepackage[skip=2pt]{caption}
\usepackage{subcaption}

\usepackage{comment}

\newcommand{\be}{\begin{equation}}
	\newcommand{\ee}{\end{equation}}
\newcommand{\beq}{\begin{equation}}
	\newcommand{\eeq}{\end{equation}}
\newcommand{\bea}{\begin{eqnarray}}
	\newcommand{\eea}{\end{eqnarray}}

\newcommand{\FTP} {\frac{\partial}{\partial \ln \eps}}

\newcommand{\eps}{{\epsilon}}

\newcommand{\MING}[2]{\textcolor{blue}{[MING: #1]}}

\title{A Worldsheet Derivation of the Black Hole Entropy}
   \author[a]{Amr Ahmadain}
   \author[b]{Ming Yang}

   \affiliation[a] {Department of Physics, Swansea University, Swansea, SA2 8PP, UK}
   \affiliation[b] {DAMTP, University of Cambridge, Cambridge CB3 0WA, UK}

   \emailAdd{amrahmadain@gmail.com}
   \emailAdd{my365@cam.ac.uk}

\abstract{We study the nonlinear sigma model (NLSM) worldsheet action describing the motion of closed bosonic strings in the target space of a two-dimensional (2D) flat cone in polar coordinates. We calculate the cylinder partition function. We first place the cylindrical worldsheet on a rectangular lattice before taking the continuum limit. We find an integer number of string configurations on the worldsheet, which we call line defects, that run from one boundary of the cylinder to the other. We insert two sources (conical defects) at each boundary and fix the two ends of the line defect by Dirichlet boundary conditions to a point $r_c$ in target space. In target space, a line defect appears as an Susskind\&Uglum-type open string ending on $r_c$. We compute the semiclassical contribution to the off-shell cylinder amplitude by saddle point approximation. The amplitude has an interesting infrared (IR) divergence structure that depends on the given range of the cone angle. We then compute the entropy by varying the cone angle. In a particular renormalization scheme that relates the ultraviolet (UV) and to the infrared (IR) limits of the modulus integral, we find the entropy to be free of IR divergences but linearly dependent on the radial cutoff. We argue that our calculation provides a well-defined state on a constant Euclidean-time slice directly from the string worldsheet. We also study the 2D flat cone NLSM  without discretization. We compute the entropy from the off-shell stationary action and show it is finite in each winding sector $W$ with a maximum at $r_c=\sqrt{\alpha'}/|W|$. After summing over all winding sectors, it still has a finite maximum in the UV limit but for $r_c >0$.}

\title{Strings at the Tip of the Cone and Black Hole Entropy From the Worldsheet: Part I}

\begin{document}

\maketitle


\section{Introduction}

In gravity, it was shown by Bekenstein and Hawking \cite{Bek,Hawk}, that the classical black hole entropy is proportional to the area of the black hole horizon \footnote{in units where $h=c=k=1$.}
\begin{equation}
 S_{\text{BH}}= \frac{A}{4 G_N},   
\end{equation}
In general relativity, one can derive $S_{\text{BH}}$,  either on-shell or off-shell. On-shell, the temperature and mass $M(\beta)$ of a black hole are not independent. In Rindler space, varying the size of thermal Rindler circle (at infinity) away from its on-shell value $\beta_{\text{R}} = 2\pi$ induces a shift in $M(\beta)$ such that the background remains on-shell. In this case, the entire contribution to $S_{\text{BH}}$ comes from the Gibbons-Hawking-York (GHY) boundary term \cite{York:1972sj,Gibbons:1976ue}.
\begin{equation}
S_{\mathrm{BH}}=\left(1-\beta \partial_\beta\right)(-\beta F)=\beta^2 \partial_\beta F(\beta) .
\end{equation}
In the off-shell method, on the other hand, the first-order $\beta$ variation is independent of $M$ in such a way that the geometry does not backreact to the change in $\beta$. Necessarily, this introduces a conical singularity at the origin of Rindler space (the black hole horizon). In this case, the Gibbons-Hawking-York term is linear in $\beta$ and therefore, does not contribute to the entropy.

Over the last three decades, there have been several attempts to derive $S_{\text{BH}}$ in string theory, the most famous of which is that by Strominger and Vafa \cite{StromingerVafa:1996}, where, for an extremal black hole, they were able to explicitly count BPS states. Susskind and Uglum \cite{SU-1994}, within the framework of induced gravity \cite{Sakharov-induced:1967,Visser-induced:2002}, and using Tseytlin's off-shell prescription \cite{TSEYTLINMobiusInfinitySubtraction1988}, computed the bulk sphere partition function and obtained the low-energy effective action (the Einstein-Hilbert term) to leading order in $\alpha'$ from the bosonic two-dimensional NLSM. Their calculation was not explicit, however, and relied entirely on the validity of Tseytlin's off-shell $\FTP$ (\textbf{T1}) prescription, which was not at all obvious. Tseytlin's \textbf{T1} prescription was later studied and verified in \cite{Ahmadain:2022tew} along with the more powerful \textbf{T2} prescription \cite{TseytlinTachyonEA2001,TseytlinSigmaModelEATachyons2001} which, in addition to the closed string massless modes, also accounts for tachyon insertions on the worldsheet. See \cite{Ahmadain:2024hdp} for a non-perturbative extension of the \textbf{T2} prescription. See also \cite{Ahmadain:2024hgd,Ahmadain:2024uom,Ahmadain:2024uyo} for an off-shell derivation of the dilaton and Gibbons-Hawking-York boundary terms from the worldsheet.

Susskind and Uglum further speculated in \cite{SU-1994} that explicit mircrostate counting is possible even for black holes far from the extremal limit, specifically, in Rindler spacetime (i.e., the infinite mass limit of a Schwarzschild black hole). This claim directly implies that $S_{\text{BH}}$ has a statistical interpretation with a Hamiltonian that acts on a one-sided and thus, a factorizable Hilbert space. In other words, their claim amounts to the statement that the classical black hole entropy is the von Neumann entropy of a reduced density matrix for an observer in the right Rindler wedge. They speculated that the sphere diagram (which goes like $\sim \frac{1}{g_s^2}$) can be viewed, from the perspective of a Rindler observer, as an open string state on a constant Euclidean-time (cone) slice, with its endpoints pinned to the codimension-2 horizon. In this picture, $S_{\text{BH}} / A=O\left(1 / g_s^2\right)=O\left(1 / G_N\right)$. To date, this open string picture remains purely speculative except in two-dimensional topological strings \cite{DW-Ebranes:2016}, where it has been made precise.\footnote{See also \cite{Takayanagi-GravityHaywardTerms:2019} for another on-shell attempt to make sense of these open strings.}


In this paper, we make an attempt---in what we consider a toy model---to explain how these horizon-localized open strings can possible arise on the worldsheet. Specifically, we study the nonlinear sigma model of strings moving in the off-shell background of two-dimensional (2D) flat cone with arbitrary cone angle $(\beta)$ (not to be confused with the thermal circle). By placing the cylindrical worldsheet on a lattice, we find that these degrees of freedom appear as particular worldsheet configurations, concretely, as line defects extending from one end of the cylinder to the other. The two cylinder boundaries are mapped to a single point $r_c$ in target space and held fixed there by Dirichlet boundary conditions. In target space, these line defects appear as open strings with their endpoints pinned to $r_c$.  There is an integer number of such line defects, equally spaced along the $(\sigma)$-direction of the worldsheet; they label distinct winding sectors and all live on a constant Euclidean-time slice, as we argue in Section \ref{sec:constant_time_slice}. 

After computing the cylinder amplitude with zero and with one defect as a function of the cone angle $\beta$, we compute the tip-localized entropy by varying $\beta$. We call the entropy tip-localized because they arise from these line defects which appear as open strings. We show that the entropy is free of UV (on the worldsheet) and IR divergences in a renormalization scheme that relates the UV and IR limits of the cylinder amplitude. However, it depends linearly on the value of $r_c$.

But let us be clear. We are \textit{not} claiming that we are able show the classical black hole entropy has origins in explicit microstate counting, as first speculated by Susskind and Uglum; rather we are merely presenting an attempt to understand the worldsheet nature of these exotic open strings by studying the 2D NLSM of bosonic strings in the background of a 2D cone with \textit{arbitrary} conical deficit. We are aware that, as explained in \cite{ahmadain2022off}, that it is not even realistic to assume that the string crosses the horizon only at two points. Indeed, without a proper regulator, a closed string configuration crosses the horizon an infinite number of times. We are not doing that in this paper.

\medskip

\noindent\textbf{Related Work.}
The literature on black hole entropy is vast and spans several decades, making it impossible to survey comprehensively within the scope of this paper. Our intention is therefore not to provide an exhaustive review, but rather to cite those that we believe are possibly related to the work done in this paper and from which we have benefited througout the course of this project.

A successful \textit{on-shell} approach to computing black hole entropy in string theory is the orbifold method
\cite{Dabholkar-Orbifold1994,StromingerLowe-Orbifold-1994} of the exact CFT $\mathbb{R}^2/\mathbb{Z}_N$, where $N$ is an odd integer\footnote{In \cite{Takayanagi-EE-StingTheory-2015}, an attempt was made to compute the entropy in Rindler space for $N$ even.}; the orbifold fixed point corresponds to the tip of the cone. This construction is inspired by the replica trick \cite{Callan:1994py,Calabrese:2009qy} in local quantum field theories \cite{Callan:1994py,Calabrese:2009qy}, where one computes entanglement entropy by evaluating the path integral on an $N$-sheeted cover of flat Minkowski spacetime and then analytically continuing in $N$. In string theory, however, the background is a cone with opening angle $2\pi/N$ rather than $2\pi N$. Most of the conceptual and technical subtleties of this method lie in the analytic continuation step. See for example, \cite{WittenOpenStringsRindler2018} and \cite{ Dabholkar:2023ows,Dabholkar:2023yqc, Dabholkar:2023tzd, Dabholkar:2024neq, Dabholkar:2025hri} for more recent results. A well-known limitation of the orbifold method is that it computes only the one-loop correction to the entropy because the tree-level contribution vanishes on-shell and it is not known how to derive a boundary term associated with the orbifold singularity that could contribute at order $(g_s^{-2})$. See \cite{Ahmadain:2022eso} for an elaborate discussion of orbifold method.

Over the past three decades, there has been extensive work on the $SL(2,\mathbb{R})$ Wess-Zumino-Witten (WZW) model describing strings in $AdS_3$ and its $SL(2,\mathbb{R})/U(1)$ coset theory, the 2D black hole and its Euclidean version (the cigar) \cite{Witten-BH:1991, Elitzur:1990ubs,Mandal:1991, Gibbons:1992rh}. In what follows, we briefly review this material, starting with the cigar and then $AdS_3$.

In \cite{KT:2001}, an attempt was made to derive the \textit{on-shell} sphere partition function (and entropy) for the cigar from a boundary term. This boundary term was not explicitly derived from the worldsheet, but rather written down using the usual logic that the full (bulk and boundary) tree-level effective action should have a well-defined variational principle. Although the black hole entropy was found in agreement with the result in \cite{Gibbons:1992rh}, the sphere partition function was found to vanish in contradiction to the result from matrix quantum mechanics by \cite{KKK:2000}.\footnote{It was nonzero, however, in a particular regularization where the dilaton field was mapped to the Liouville field.}  See also \cite{Ahmadain:2022gfw} for a microscopic derivation of the entropy in sine-Liouville theory.\footnote{See \cite{kraus2002strings} for an attempt to derive the boundary on-shell action from the string worldsheet. See \cite{Hartnoll:2015fca} for calculating entanglement entropy in two dimensions from the target space and \cite{Naseer2020} for an attempt to calculate entanglement entropy from closed string field theory action.}

The string-scale ($\ell_s$) structure of the Euclidean black hole horizon has also received a lot of attention. The physics of the winding tachyon condensate (the Fateev, Zamolodchikov and Zamolodchikov (FZZ) duality to sine-Louville theory \cite{FZZ,KKK:2000}) and phase transition at the tip of the cigar \cite{Giveon:2012kp,Giveon:2013ica,Giveon:2015cma,Giveon:2016dxe}, the black hole interior \cite{Itzhaki:2018glf,Itzhaki:2018rld,Giveon:2019gfk,Giveon:2019twx}, and the physics of the thermal scalar and random walks at the tip of the cigar \cite{Mertens-RandomWalk:2013,Mertens:2013pza,Mertens:2014dia} (see \cite{Mertens:Thesis:2015} for a comprehensive overview of some of these these topics.) 

The Horowitz-Polchinski solution is a self-gravitating gas of oscillating strings in a low number of dimensions that involves a local winding condensate \cite{HP2:1997}. The Horowitz-Polchinski solution is valid in the regime where $\beta-\beta_H \ll l_s$ \cite{HP2:1997}, where $\beta$ the radius of the asymptotic radius of the thermal circle and $\beta_H$ the radius at the Hagedorn temperature. The most surprising feature of the Horowitz-Polchinski solution is that it has a non-zero \textit{classical} entropy, expected of the Schwarzschild black hole solution. The Horowitz-Polchinski solution has recently been studied \cite{Chen-LargeD:2021} in the large-D expansion \cite{Emparan:2013xia} where it was argued that the temperature of a large-D black hole can go beyond the Hagedorn temperature and transition into gas of strings. It was also observed that the winding condensate contributes to only part of the total entropy of the cigar, but the contribution can dominate as $k\rightarrow3$\footnote{Large $k$ is the low-energy (weak-curvature) regime where gravity is valid while small $k$ describes the stringy regime.}. This was followed by a study of the transition between black holes and highly-excited strings in the Horowitz-Polchinski solution \cite{Chen:2021dsw} for bosonic, Type-II and heterotic string theory. On the other hand, in \cite{Brustein:2021qkj}, it was shown that the \textit{entire} black hole entropy can be accounted for by the entropy of the winding condensate. By solving a reduced version of the Horowitz-Polchinski equations of motion, the authors found a critical solution which takes the form of a cigar geometry with a puncture at the tip. They further argued that this puncture can remove the obstruction found in the Witten index for type-II strings \cite{Chen:2021dsw}.

Equally importantly, the exact worldsheet CFT describing strings on $\mathrm{AdS}_3$ with NS-NS flux \cite{deBoer:1998gyt,giveon1998comments,kutasov1999more,maldacena2001stringsI,maldacena2001stringsII,maldacena2002stringsIII,Troost:2002wk} has provided a successful laboratory for string theory in curved spacetime. In this background, the $\mathrm{SL}(2, \mathbb{R})$ WZW model made it possible to test various aspects of the gauge-gravity duality \cite{Maldacena:1997re,Witten:1998qj} in an exactly solvable worldsheet theory \cite{Eberhardt-WS-Dual:2018,Eberhardt-SymmetricOrbifold:2019l,Eberhardt-PF-tensionless-String:2020,Eberhardt:2021vsx}. Using path integral techniques, the thermal partition function was computed in \cite{Hanany:2002ev} (see also \cite{Ashok:2020dnc} and \cite{Ferko:2024uxi} for recent results). An attempt to compute the sphere partition function was made in \cite{eberhardt2023holographic} (but see also \cite{Mahajan-sphere:2021}).

Recently, the authors of \cite{Halder:2023adw} evaluated the leading tree-level contribution to entropy for a BTZ black hole \cite{Berkooz:2007fe} in terms of correlation functions of the winding-operator. They defined a Lewkowycz-Maldacena \cite{LM2013} stringy replica trick approach for computing entanglement entropy in terms of a winding condensate, and ultimately reduced the computation to the one-point function of a non-local area operator\cite{Halder:2023adw}. In \cite{Halder:2024gwe}, they used this stringy replica trick to compute the thermal entropy of the cigar in the large-D expansion and found it agrees with the results in \cite{Callan:1988hs}, in the $\alpha' \rightarrow 0$ limit.

\subsection{Strategy and main results}
Here, we outline our strategy for calculating the entropy and state our results.

\medskip

We place the worldsheet NLSM on a cylindrical lattice of length $\ell_\tau$. The NLSM has two background (target space) fields: $r(\tau, \sigma)$, non-compact radial boson and $\theta\sim\theta+2\pi$ (compact Euclidean thermal circle). We denote the cone angle by $\beta$ and define $\beta =1$ to be the Rindler space. The UV cutoff (lattice spacing) is $\eps$.

We perform a Hubbard–Stratonovich transformation to linearize the lattice interaction; this introduces a quantized auxiliary current $J(\sigma, \tau)$. After integrating out the Euclidean time $\theta$-circle\footnote{In doing this, we were inspired by these two papers \cite{Inomata:2011pxd,Inomata:2011md}}, The path integral takes the form of a sum over string worldsheet configurations with a divergence-free current $\nabla\!\cdot J=0$ (see Section \ref{sec:Calc} for a proper definition of all quantities in $K_{\mathrm{cyl}}$)
\be
K_{\mathrm{cyl}}\left(\beta, \ell_\tau, r_c\right)=\sum_{\nabla \cdot J=0} \int\left[\prod_{n \in V} d r_n\right]\left[\prod_{n \in V} \frac{1}{4 \pi \beta r_n}\right] \exp \left\{-\sum_{\langle m, n\rangle \in L}\left[\frac{1}{4\pi\alpha'}\left(r_m-r_n\right)^2+\frac{\pi \alpha'J_{m n}^2}{\beta^2 r_m r_n}\right]\right\}.
\ee

\begin{figure}
    \centering
    \includegraphics[width=0.8\linewidth]{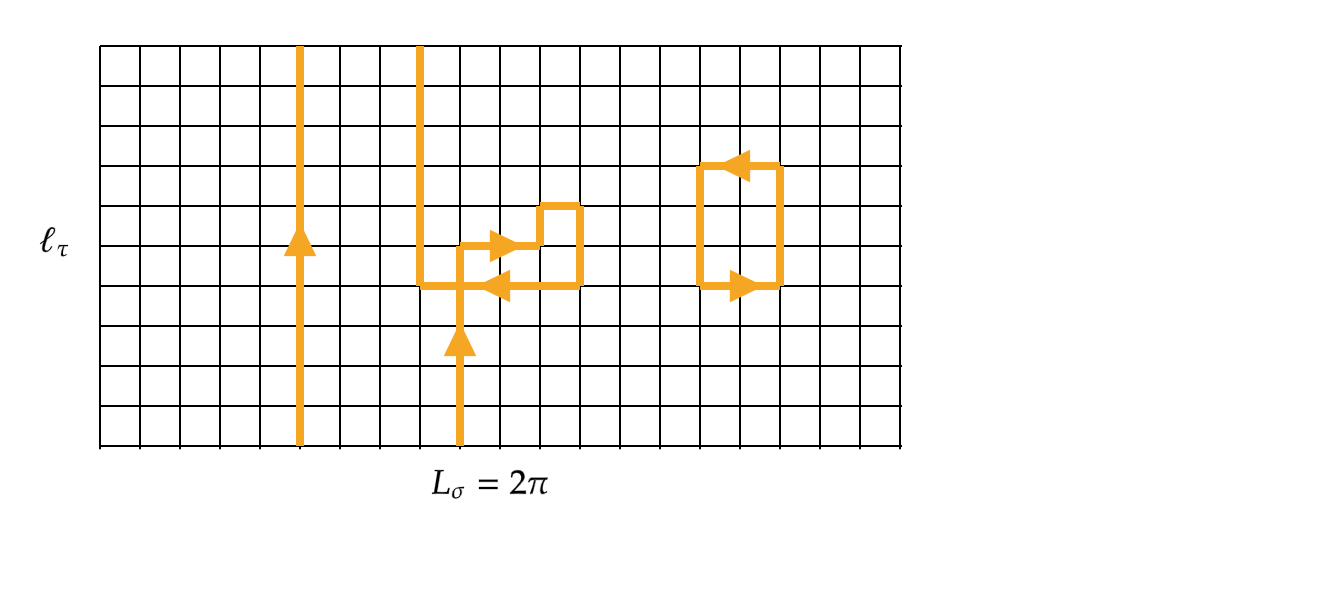}
    \caption{Classes of string configurations on the worldsheet. The straight line defects are energetically favorable as they cross the least number of links on the lattice, compared to the other two classes of configurations. The string configuration in the middle is a small fluctuation about the straight line defect, which is why, the well approximated by the former.}
    \label{fig:config}
\end{figure}

Broadly speaking, the lattice has two classes of worldsheet configurations: (1) bulk loops (closed lines entirely in the interior of the worldsheet) and (2) lines that extend along the length of the cylinder from one boundary to another, i.e. run between the two conical defects. See fig. \ref{fig:config}. The  former satisfies $\nabla\!\cdot J=0$ while the latter violates it at the boundaries. This can be understood as inserting two sources, one at each boundary, where a flux line terminates. As a result, there is \textit{net} outflow, $+N_{\text{def}}$ and inflow, $-N_{\text{def}}$ and thus, $\nabla \cdot J$ is not \textit{locally} conserved.

However, because of the weight $\exp(-J^2/r^2)$, loop configurations are energetically suppressed, and thus the dominant contribution comes from straight defect-to-defect lines. In fact, there are other worldsheet configurations that would have been necessarily taken into account if it were not for the fact that the worldsheet topology is fixed to be a cylinder. These additional configurations can potentially probe critical phase transition on the worldsheet, but we will defer the discussion of this point to Section \ref{sec:discussion}.

The NLSM action now is the sum of two terms: (i) the action of a free non-compact boson $r$ and (ii) the sum of an inverse-square potential term localized on the defect lines along the $\sigma$-direction with the two endpoints of each line defect fixed at $r=r_c$\footnote{(which perhaps can be thought of as the location of a stretched horizon \cite{Sen:2004dp}).} by Dirichlet boundary conditions. Concretely, $K_{\mathrm{cyl}}\left(\beta, \ell_\tau, r_c\right)$ is the sum of (a gas of) ${N_{\text{def}}}$ distinct line defect (or flux line) configurations ${\gamma_k(s)}$ in the background of an inverse-square radial potential 
\begin{equation}
K_{\mathrm{cyl}}\left(\beta, \ell_\tau, r_c\right)=\sum_{\gamma_k(s)}\int_{\substack{r(0, \sigma) = r_c \\ r(\ell_\tau, \sigma) = r_c}} \left[\mathcal{D}r\right]\left[{\frac{1}{4\pi r\beta}}\right]\exp(- \frac{1}{4 \pi \alpha'} \int d^2z \, \partial r\partial r + \frac{\pi\alpha^{\prime}}{ \beta^2 \epsilon} \sum_{k=1}^{N_{\text{def}}} \int_0^1 d s \, \frac{\left|\dot{z}_k(s)\right|}{r\left(z_k(s)\right)^2}),
\end{equation}
where $s$ is the intrinsic length parameter of a given flux line and $\dot{z}_k(s) = \frac{dz(s)}{d\tau}$.

\begin{figure}
  \centering
  \includegraphics[width=1.0\linewidth]{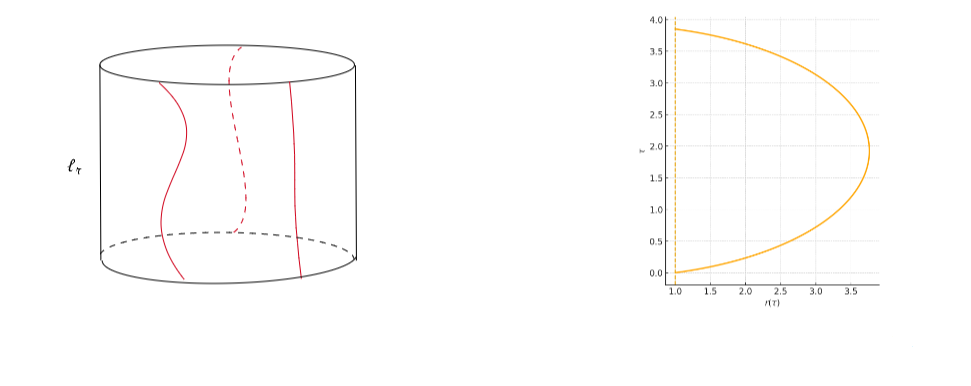}
  \vspace{-1cm}
  \caption{Left: The cylindrical worldsheet with three defect lines ($N_{\text{def}} =3$) extending from one side to another. Right: The target space radial trajectory $r(\tau)$ for one defect line. It appears as an open string on a constant Euclidean-time ($\theta$) slice with its two endpoints pinned to $r_c$. The trajectory starts at $r_c$, rises to $r$ at $\ell_{\tau}/2$ before it folds back to $r_c$ again. The Dirichlet boundary conditions at $r_c$ act as attractive sources that force the string to return to $r_c$.}
\label{fig:cylinder-open-string}
\end{figure}

The target space image of a defect line with its two endpoints pinned to $r_c$ is reminiscent of the open string in Susskind and Uglum \cite{SU-1994}; it lives on a constant Euclidean-time slice. See fig. \ref{fig:cylinder-open-string}.

\medskip

Assuming the total Hilbert space factorizes into independent sectors\footnote{$\mathcal H \;=\;\bigotimes_{N_{\text{def}}}\,\mathcal H_{N_{\text{def}}}$ with $H = \sum_{N_{\text{def}}} H_{N_{\text{def}}}$.}, the target space string amplitude takes the form 
\begin{equation}
\mathcal{Z}_T(\beta) = \prod_{N_{\text{def}}} \tilde{\mathcal{Z}}_{N_{\text{def}}}(\beta), \qquad \tilde{\mathcal{Z}}_{N_{\text{def}}}(\beta):=\operatorname{Tr} \rho^{\beta}_{N_{\text{def}}},
\end{equation}
Taking the logarithm gives
\begin{equation}
\mathcal{K}(\beta)\coloneqq\log \mathcal{Z}_T(\beta)=\sum_{N_{\text{def}}} \log\tilde{\mathcal{Z}}_{N_{\text{def}}}(\beta)
= \sum_{N_{\text{def}}} \mathcal{Z}_{N_{\text{def}}}(\beta),
\end{equation}
where $\mathcal{Z}_{N_{\text{def}}}$ is the amplitude in the $N_{\text{def}}$ sector
\begin{equation}
\mathcal{Z}_{N_{\text{def}}} = \int d \ell_{\tau} \, K_{N_{\text{def}}}(\beta,\ell_\tau; r_c),
\end{equation}
and 
\begin{equation}
K_{N_{\text{def}}}(\beta,\ell_\tau; r_c) := \left\langle r_c \right| e^{-\ell_\tau H_{N_{\text{def}}}(\beta)} \left| r_c\right\rangle,
\end{equation}
the partition function for $N_{\text{def}}$ line defects. $\mathcal{Z}_{N_{\text{def}}}$ is the quantity we calculate in this paper.

\medskip

In string theory, $\operatorname{Tr}\hat{\rho}^\beta$ is formally the exponential of the sum of contributions of connected string worldsheet amplitudes. The logarithm of this sum $\log \operatorname{Tr}\hat{\rho}^\beta$ is simply the sum of these amplitudes
\begin{equation}\label{eq:renormalzed_Z}
\widehat{\mathcal{Z}}(\beta) \coloneqq \mathcal{Z}(\beta) -\beta \mathcal{Z}(1) \coloneqq \log \operatorname{Tr} \hat{\rho}^\beta \,,
\end{equation}
where $\widehat{\mathcal{Z}}(\beta)$ is the normalized partition function and $\hat{\rho} = \frac{\rho}{\operatorname{Tr} \rho}$ is the normalized density matrix for an observer outside the horizon (at $r >0$) with $\operatorname{Tr} \hat{\rho}=1$.  

The entropy is computed by varying $\beta$ 
\begin{equation}
S=(1-\left.\beta \partial_\beta) \widehat{\mathcal{Z}}(\beta)\right|_{\beta=1} =\mathcal{Z}(1)-\mathcal{Z}^{\prime}(1) =-\operatorname{Tr}(\hat{\rho} \ln \hat{\rho}).
\end{equation}

In this paper, we calculate the $\mathcal{Z}_0$ (the zero-defect sector) amplitude and find it to be linear in $\beta$ and thus, as expected, the associated entropy vanishes upon varying the cone angle $\beta$
\begin{equation}
S_0=(1-\beta\partial_\beta)|_{\beta=1}\,\mathcal{Z}_0(\beta)=0.
\end{equation}

In the single-defect sector ($N_{\mathrm{def}}=1$), we reduce the 2D action to an interacting 1D line-defect action with a $1/r^2$ potential and impose Dirichlet boundary conditions to fix the two boundaries of the cylinder to $r_c$; we then determine its stationary point by the method of saddle point approximation. In the IR (long cylinder) limit, we find that the amplitude $\mathcal{Z}_1(\beta,\Lambda, \eps)$ has a divergence pattern that depends on the range $\beta$
\begin{equation}
\mathcal{Z}_1(\beta,\Lambda, \eps) = \int^{\Lambda}\!d\ell_\tau\, K_1(\beta, \ell_{\tau}, \eps)
\;\sim\;
\begin{cases}
\dfrac{C(\beta,\eps)}{1-\frac{1}{2\beta}}\ \Lambda^{\,1-\Delta(\beta)}, & \beta>\dfrac{1}{2}\quad(\text{power law divergence}),\\[8pt]
C(\beta,\eps)\,\ln\Lambda, & \beta=\dfrac{1}{2}\quad(\text{log divergence}),\\[8pt]
\dfrac{C(\beta,\eps)}{\Delta(\beta)-1}, & \beta<\dfrac{1}{2}\quad(\text{finite in $\Lambda$}).
\end{cases}
\end{equation}
where $\Delta(\beta) = \frac{1}{2\beta}$ is the scaling dimension of the defect line operator and $\Lambda$ is an IR cutoff.

\medskip
In a particular renormalization scheme and after applying Tseytlin's prescription \cite{TSEYTLINMobiusInfinitySubtraction1988,Ahmadain:2022tew}, we find that the entropy computed from $\mathcal{Z}_1(\beta,\Lambda, \eps)$ is free of IR divergences but depends linearly on $r_c$ 
\begin{equation}
S(r_c) =(1-\beta\,\partial_\beta) \,\mathcal{Z}_1(\beta,\Lambda,\eps)|_{\beta = 1} = \frac{r_c}{\sqrt{\pi\alpha'}}.
\end{equation}
We will attempt to explain the implications of this linear dependence on $r_c$ in Section \ref{sec:saddle_point}.

\medskip

In Section~\ref{sec:SC_saddle_Circular}, we revisit the same NLSM on the flat 2D cone but without discretizing the worldsheet or using a Hubbard–Stratonovich transformation; instead we reduce the 2D action to 1D by the ansatz $r(\tau,\sigma)=r(\tau)$, $\theta(\tau,\sigma)=W\sigma$ with $W\in\mathbb{Z}$ and Dirichlet boundary condition $r(0)=r(\ell_\tau)=r_c>0$. We compute the classical (off-shell) stationary-point action, and obtain the semiclassical partition function in the long-cylinder IR regime. We compute the entropy and show it is finite in each winding sector $W$ with a maximum at $r_c=\sqrt{\alpha'}/|W|$. The $|W|=1$ sector, therefore, has the largest contribution. Upon summing over all $|W| \ge 1$, the entropy attains a finite non-universal constant $4\log 2$ even as $r_c\to 0^{+}$
\begin{equation}
S(r_c)=2\sum_{W\in\mathbb{Z}\setminus\{0\}}S_{\text{finite}}(W,r_c)
=4\log 2\,\frac{r_c^2}{\alpha'}\sum_{W=1}^{\infty}e^{-W r_c^2/\alpha'}=4\log 2\;\frac{r_c^2}{\alpha'}\;\frac{1}{e^{\,r_c^2/\alpha'}-1}\;.
\end{equation}
We show that $S(r_c)$ converges even in the UV limit (for $\ell_\tau \ll 1$) but only for $r_c >0$. We also show that it can be expressed as the mean energy (with respect to the Bose-Einstein distribution) to add a winding mode quantum to a harmonic oscillator.

\medskip

\noindent\textbf{Paper layout.} The paper is organized as follows. In Section \ref{sec:Calc}, we introduce the NLSM on the cone and set up the lattice description of the cylindrical worldsheet. We do a Hubbard-Stratonovic transformation and obtain an effective worldsheet action. We specify the boundary conditions and write down the full cylinder path integral for $N_{\text{def}}$ line defects. In Section \ref{sec:saddle_point}, we compute the semiclassical action on a half-line by the stationary phase approximation. We then compute the cylinder amplitude in the IR limit as a function of the cone angle $\beta$ and comment on its infrared divergences which depend on the range of $\beta$. We then compute the tip-localized entropy by varying $\beta$, and show how a specific renormalization scheme can yield it finite. In Section $\ref{sec:constant_time_slice}$, we argue how integrating out the Euclidean thermal circle defines a state on a constant-Euclidean time slice of the cone. In Section~\ref{sec:SC_saddle_Circular}, we study the 2D flat cone NLSM  without discretization. We compute the entropy from the off-shell stationary action and find it to be finite in each winding sector with a maximum at $r_c=\sqrt{\alpha'}/|W|$. After summing over all winding sectors, it still has a finite maximum even in the UV limit but for $r_c >0$. Finally, in Section \ref{sec:discussion}, we comment on various aspects of our work, present some questions before we end with possible future directions.


\section{The Worldsheet on a Lattice}\label{sec:Calc}
In this section, we explain our lattice setup in detail. We explain the action for the defect line configurations, the partition function, the measure factor, and the boundary conditions. We also present the calculation for $K_0$.

\subsection{The lattice}
We take the worldsheet to be a cylinder
$\Sigma=S_\sigma^1\times\left[0, \ell_\tau\right]  $  with coordinates 
\begin{equation}
(\sigma, \tau) \in [0, 2\pi] \times [0, \ell_\tau], \quad \sigma \sim \sigma + 2\pi.
\end{equation}

Consider the NSLM action which describes the motion of closed bosonic strings in the target space of a 2D flat cone with cone angle $\beta < 1$\footnote{In a conical spacetime with surplus (deficit) angle $\delta = 2\pi \pm \beta_{\text{cone}}$, the local inverse temperature $\beta(r)$ is $\beta_{\text{cone}}\, r$ which means an observer at fixed $r$ experiences a larger (smaller) local temperature, respectively.}
\begin{equation}
I[r, \theta]=\frac{1}{4 \pi \alpha^{\prime}} \int_{\Sigma}\left[(\partial r)^2+\beta^2 r^2(\partial \theta)^2\right], \qquad r\ge0,\quad \theta\sim\theta+2\pi,
\end{equation}

\begin{figure}
    \centering
    \includegraphics[width=0.8\linewidth]{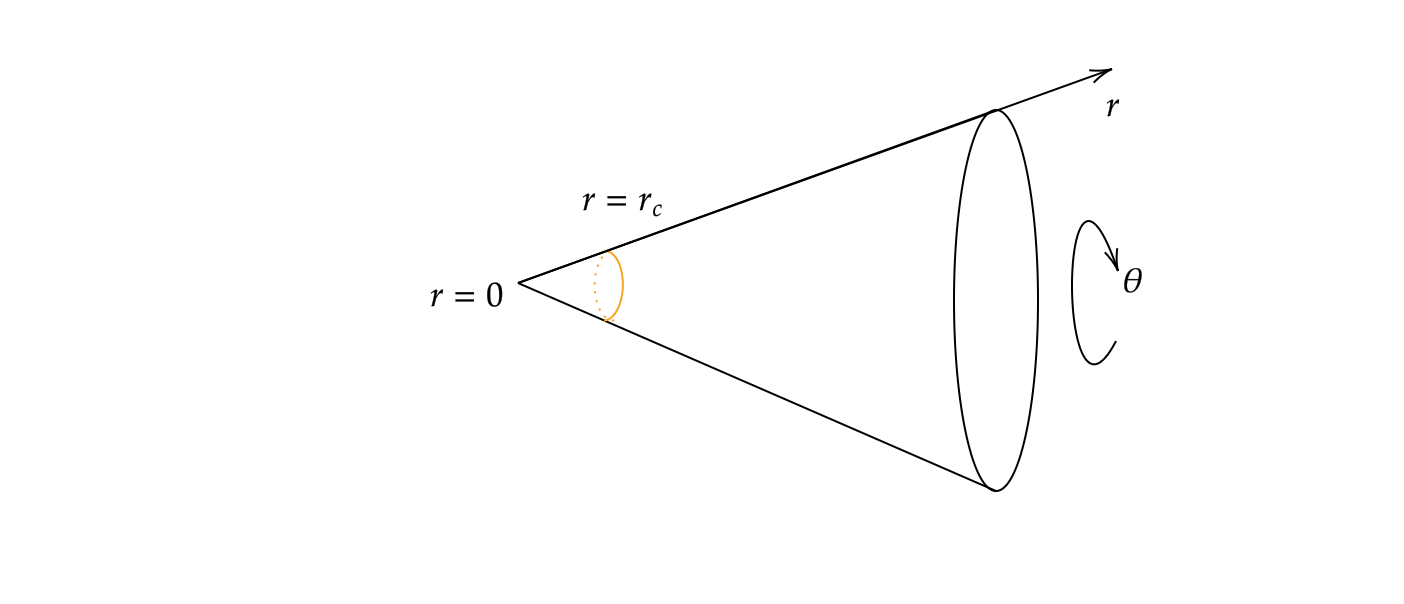}
    \caption{Our target space is a flat 2D cone. $r_c$ is where we fix the two boundaries of the cylinder in target space. It also regularizes the conical singularity at $r=0$.}
    \label{fig:cone}
\end{figure}

The cylinder partition function is given by
\begin{equation}
K_{\mathrm{cyl}}\left(\beta, \ell_\tau, r_c\right)=\int_{\substack{r(0, \sigma) = r_c \\ r(\ell_\tau, \sigma) = r_c}}\left[\prod_{n}\left(\int_0^{\infty} d r_n\right)\left(\int_0^{2 \pi} d \theta_n\right)\right]\left[\prod_{n} r_n\right] \exp (-I[r, \theta]),
\end{equation}
where $r_c$ is cutoff radius in target space to be defined precisely later. 

The cylinder amplitude is the integral over the cylinder length $\ell_{\tau}$ of $K_{\text{cyl}} (\beta)$
\begin{equation}
\mathcal{Z}_{\text{cyl}} (\beta, r_c) = \int_0^\infty d \ell_\tau \, K_{\mathrm{cyl}}\left(\beta, \ell_\tau, r_c\right).
\end{equation}

We place the worldsheet theory on a flat rectangular lattice. A site $n=(a, b)$, where $a \in\left\{0,1, \ldots, N_\sigma-1\right\}$ labels the $\sigma$ direction and $b \in\left\{0,1, \ldots, N_\tau\right\}$ labels the $\tau$ direction. In the $\sigma$-direction, $a$ is taken $\bmod \, N_\sigma$ (wraps around horizontally) while in the open $\tau$-direction, $b=0$ is the bottom circle, $b=N_\tau$ is the top circle. Unit steps are denoted $\hat{\sigma}=(1,0)$ and $\hat{\tau}=(0,1)$. 

The discrete worldsheet action takes the form
\begin{equation}\label{eq:Discrete_WS_action}
I_{\text{w.s.}}[r,\theta]={\frac{1}{4\pi\alpha'}}\sum_{<m,n>} (r_m-r_n)^2+r_mr_n\beta^2(\theta_m-\theta_n)^2.
\end{equation}
where $<m,n>$ denotes the nearest neighbor.

In terms of \eqref{eq:Discrete_WS_action}, the cylinder partition function reads 
\begin{align}\label{eq:K_lattice}
\begin{split}
K_{\mathrm{cyl}}\left(\beta, \ell_\tau, r_c\right)=&\int\left[\prod_{n \in V}\left(\int_0^{\infty} d r_n\right)\,[r_n]\,\left(\int_0^{2 \pi} d \theta_n\right)\right] \,\prod_{m \in \partial V} \delta\left(r_n-r_b\right) \\ 
&\exp \left\{-\sum_{\langle m, n\rangle \in L}\left[\left(r_m-r_n\right)^2+\beta^2 r_m r_n\left(\theta_m-\theta_n\right)^2\right]\right\},
\end{split}
\end{align}
where $V$ is set of vertices (sites) of the lattice, $\partial V$ denotes the two boundaries of the cylinder, and $L=\{\langle m, n\rangle\}$ is set of nearest-neighbor links.\footnote{Using the term $[r_n]$ in the measure factor, $K_{\mathrm{cyl}}\left(\beta, \ell_\tau, r_c\right)$ can be expressed as 
$$\label{eq:K_lattice1}
K_{\mathrm{cyl}}\left(\beta, \ell_\tau, r_c\right)=\int\left[\prod_{n \in V}\left(\int_0^{\infty} d r_n\right)\,\,\left(\int_0^{2 \pi} d \theta_n\right)\right] \exp \left\{-\left(\sum_{\langle m, n\rangle \in L}\left[\left(r_m-r_n\right)^2+\beta^2 r_m r_n\left(\theta_m-\theta_n\right)^2\right]-\sum_{n\in V}\log r_n\right)\right\}.
$$.}

Using the Hubbard-Stratonovic transformation, \eqref{eq:K_lattice} becomes
\begin{align}\label{eq:K_cyl_HS}
K_{\mathrm{cyl}}\left(\beta, \ell_\tau, r_c\right)= & \int\left[\prod_{n \in V}\left(\int_0^{\infty} d r_n\right)\right]\left[\prod_{\langle m, n\rangle \in L} \sqrt{\frac{1}{4 \pi \beta^2 r_m r_n}}\right]\left[\prod_{n \in V}\left(\beta r_n\right)\right] \prod_{m \in \partial V} \delta\left(r_n-r_b\right) \nonumber \\
& \times\left[\prod_{n \in V}\left(\int_0^{2 \pi} d \theta_n\right)\right]\left[\prod_{\langle m, n\rangle \in E}\left(\int_{-\infty}^{\infty} d J_{m n}\right)\right] \\ \nonumber
& \times \exp \left(-\sum_{\langle m, n\rangle \in L}\,\frac{1}{4\pi\alpha'}\left(r_m-r_n\right)^2+\sum_{\langle m, n\rangle \in L} \frac{\pi \alpha' J_{m n}^2}{ \beta^2 r_m r_n}+\sum_{\langle m, n\rangle \in L} i J_{m n}\left(\theta_m-\theta_n\right)\right)\,.
\end{align}
Here  $J_{m, n} \in \mathbb{R}$  is the oriented link current from site  $m$ to site  $n$ with antisymmetry $J_{m, n}=-J_{n, m}$.

Because $\theta_n \sim \theta_n+2 \pi$ at every site, any link term can only depend on the relative phase $e^{i\left(\theta_m-\theta_n\right)}$; equivalently, the link dependence must be $2 \pi$-periodic in $\theta_m-\theta_n$. The unique Fourier basis for such periodic functions is $\left\{e^{i J\left(\theta_m-\theta_n\right)}\right\}_{J \in \mathbb{Z}}$. Hence, we take $J_{m, n} \in \mathbb{Z}$ so that each link weight is a superposition of these harmonics.\footnote{One concrete way to realize this is to take the Villain link weight $\sum_{w \in \mathbb{Z}} \exp {-\beta^2 r_m r_n\left(\theta_m-\theta_n-2 \pi w\right)^2}$ in $K_{\text {cyl }}$. A Poisson resummation gives the exact character expansion $\propto \sum_{J \in \mathbb{Z}} e^{-J^2 /\left(4 \beta^2 r_m r_n\right)}. e^{i J\left(\theta_m-\theta_n\right)}$, making $J_{m, n}$ integer and yielding the Gaussian cost $J_{m, n}^2 /\left(4 \beta^2 r_m r_n\right)$ that appears in the exponent.}


\medskip

With this choice, we may insert the identity
\be
1=\delta_{\mathbb{Z}}\left(J_{m n}\right)=\lim _{N \rightarrow \infty} \frac{1}{2 N+1} \sum_{k=-N}^N e^{i 2 \pi k J_{m n}},
\ee
thereby projecting $J_{m n}$ onto integer values
\be
e^{i J_{m n}\left(\theta_m-\theta_n\right)}=\left(e^{i J_{m n}\left(\theta_m-\theta_n\right)}\right) \delta_{\mathbb{Z}}\left(J_{m n}\right).
\ee
Summing over all nearest-neighbor links, we then get
\be
-\sum_{\langle m, n\rangle \in L} i J_{m n}\left(\theta_m-\theta_n\right)=-i \sum_{m \in V} \theta_m\left(\sum_{n \in \backslash \operatorname{nbr}(m)} s_{m n} J_{m n}\right)=-i \sum_{m \in V} \theta_m(\nabla \cdot J)_m,
\ee
where the divergence $(\nabla \cdot J)_m$ is defined as
\begin{equation}
(\nabla \cdot J)_m \equiv J_{m, m+\hat{\sigma}}-J_{m-\hat{\sigma}, m}+J_{m, m+\hat{\tau}}-J_{m-\hat{\tau}, m}.
\end{equation}

Now we can integrate out $\theta$ in \eqref{eq:K_cyl_HS} 
\be \label{eq:theta_integral}
\int_0^{2 \pi} d \theta_m \, e^{-i \theta_m(\nabla \cdot J)_m}=2 \pi \delta\left((\nabla \cdot J)_m\right).
\ee
Taking the product over sites gives
\be
\int\left[d \theta_n\right] \exp \left(-\sum_{\langle m, n\rangle} i J_{m n}\left(\theta_m-\theta_n\right)\right)=(2 \pi)^{|V|} \prod_{m \in V} \delta\left((\nabla \cdot J)_m\right).
\ee
The $\theta$-integral enforces divergence-free condition in the \textit{bulk} (away from the two boundaries of the cylinder)
\be
(\nabla \cdot J)_m=0 \quad \text { for every interior site } m.
\ee
This is nothing but Gauss law on the lattice. 

\medskip

On the cylinder, the \textit{net} vertical flux $W$ through any constant-$\tau$ circle is 
\be
W_b=\sum_a J_{(a, b) \rightarrow(a, b+1)}.
\ee
Physically, $W \in \mathbb{Z}$ is the winding sector carried by the defect lines. Summing the divergence over all sites in the fixed row $b$ gives
\be
\sum_a(\nabla \cdot J)_{(a, b)}=\sum_a\left[J_{(a, b) \rightarrow(a+1, b)}-J_{(a-1, b) \rightarrow(a, b)}\right]+\sum_a\left[J_{(a, b) \rightarrow(a, b+1)}-J_{(a, b-1) \rightarrow(a, b)}\right] .
\ee
The first term vanishes on a $\sigma$-circle. The second term is $W_b-W_{b-1}$. Thus in the bulk, the winding number does not change
\be
\sum_a(\nabla \cdot J)_{(a, b)}=W_b-W_{b-1} =0 .
\ee

At the two boundary sites, however, we insert two sources, one at $n_{-}$ (the boundary circle at $\tau=0$) and one at $n_{+}$ (the boundary circle at $\left.\tau=\ell_\tau\right)$. There is \textit{net} outflow $+W$ and inflow $-W$ and thus, $\nabla \cdot J$ is not \textit{locally} conserved
\be
\nabla \cdot J=W\left(\delta_{m, n_{+}}-\delta_{m, n}\right),
\ee
and the winding number changes
\begin{equation}
+\sum_{m \in \partial \Sigma \,(\tau=0)}(\nabla \cdot J)_m=+W, \qquad -\sum_{m \in \partial \Sigma \,(\tau=\ell_{\tau})}(\nabla \cdot J)_m=-W.
\end{equation}
\textit{Globally}, therefore, the charge $W$ is conserved.


With these assumptions, the cylinder partition function $K_{\mathrm{cyl}}\left(\beta, \ell_\tau, r_c\right)$ is a sum of divergence-free configurations
\begin{align}\label{eq:K_cyl_J^2}
K_{\mathrm{cyl}}\left(\beta, \ell_\tau, r_c\right)=&\sum_{\nabla \cdot J=0} \int\left[\prod_{n \in V} d r_n\right]\left[\prod_{n \in V} \frac{1}{4 \pi \beta r_n}\right] \prod_{m \in \partial V} \delta\left(r_n-r_b\right) \nonumber \\
&\exp \left\{-\sum_{\langle m, n\rangle \in L}\left[\frac{1}{4\pi\alpha'}\left(r_m-r_n\right)^2+\frac{\pi \alpha'J_{m n}^2}{\beta^2 r_m r_n}\right]\right\}.
\end{align}
There are two types of divergence-free configurations (shown in fig. \ref{fig:config}) in $K_{\mathrm{cyl}}\left(\beta, \ell_\tau, r_c\right)$ with total vertical flux $W$: (1) $N_{\text{def}} = |W|$ open defect lines running from the bottom source to the top sink (carrying net flux), and (2) a number of closed loops in the interior, contributing no net flux. But which $J$ configurations dominate $K_{\text {cyl}}$?

Let $w_{m n}\coloneqq\frac{1}{4 \beta^2 r_m r_n}$. Consider any admissible configuration $J$ with $\left|J_{m n}\right|= k \geq 2$ on some link $\langle m, n\rangle$. Because the cost is quadratic and link-wise additive
\begin{equation}
w_{m n}\, k^2=w_{m n}(1+(k-1))^2 \geq w_{m n} + w_{m n}\,(k-1)^2.
\end{equation}
If $k$-unit flux (defect) line cuts through $k$ disjoint links $\left\{\ell_i\right\}$, then each carries one unit, $J_{\ell_i}= \pm 1$. Since every $w_{\ell_i} \leq \max \left\{w_{\text {local }}\right\}$, we get
\begin{equation}
\sum_{i=1}^k w_{\ell_i} \leq k \, w_{m n} < w_{m n}\, k^2.
\end{equation}
Therefore, these line configurations with $\left|J_{m n}\right| \in\{0,1\}$ strictly decreases the energy cost. Simply put, for fixed total flux across any cut, minimizing $\sum w_{\ell} x_{\ell}^2$ with $x_{\ell} \in \mathbb{Z}_{\geq 0}$ uses as many distinct links as possible. These line defect configurations thus represent the dominant contributions to the path integral. 

On the other hand, a closed loop $\mathcal{C}$ in the bulk of the lattice with unit flux costs
\begin{equation}
\Delta I[\mathcal{C}]=\sum_{\langle m, n\rangle \in \mathcal C} w_{m n}( \pm 1)^2=\sum_{\langle m, n\rangle \in \mathcal C} \frac{1}{4 \beta^2 r_m r_n}>0.  
\end{equation}
Consequently, any configuration containing additional closed loops is strictly more costly than the same configuration with those loops removed. Closed loops are energetically unfavorable and therefore, suppressed.

Now, $K_{\mathrm{cyl}}\left(\beta, \ell_\tau, r_c\right)$ can be expressed as the sum of ${N_{\text{def}}}$ distinct flux line configurations ${\gamma_k(s)}$ in an inverse radial potential background
\begin{equation}\label{eq:Kcyl_line_configs}
K_{\mathrm{cyl}}\left(\beta, \ell_\tau, r_c\right)=\sum_{\gamma_k(s)}\int_{\substack{r(0, \sigma) = r_c \\ r(\ell_\tau, \sigma) = r_c}} \left[\mathcal{D}r\right]\left[{\frac{1}{4\pi r\beta}}\right]\exp(- \frac{1}{4 \pi \alpha'} \int d^2z \, \partial r\partial r + \frac{\pi\alpha^{\prime}}{ \beta^2 \epsilon} \sum_{k=1}^{N_{\text{def}}} \int_0^1 d s \, \frac{\left|\dot{z}_k(s)\right|}{r\left(z_k(s)\right)^2}),
\end{equation}
where $s$ is the intrinsic distance parameter of a given flux line and $\dot{z}_k(s) = \frac{dz(s)}{d\tau}$. We have restored the lattice spacing $\epsilon$ by carefully keeping tracks of the number of sums in the integral of $s$.

Let $\widehat{O}_{\text{defect}}[r ;\{\gamma_k\}] =  \frac{\pi \alpha^{\prime}}{\beta^2 \epsilon} \sum_{k=1}^{N_{\text{def}}} \int_0^1 d s \, \frac{\left|\dot{z}_k(s)\right|}{r\left(z_k(s)\right)^2}$ be the operator for ${N_{\text{def}}}$ line insertions. For any worldsheet point $z^{\prime}=\left(\tau^{\prime}, \sigma^{\prime}\right)$, we have the following identity
\begin{equation}
 1=\int_0^{\ell_\tau} d \tau^{\prime} \int_0^{2 \pi} d \sigma^{\prime} \delta^{(2)}\left(z^{\prime}-z_k(s)\right),  
\end{equation}
Then $\frac{\left|\dot{z}_k(s)\right|}{r\left(z_k(s)\right)^2}$ can be written as
\begin{equation}
\frac{\left|\dot{z}_k(s)\right|}{r\left(z_k(s)\right)^2}=\int_0^{\ell_\tau} d \tau^{\prime} \int_0^{2 \pi} d \sigma^{\prime} \frac{\left|\dot{z}_k(s)\right|}{r\left(\tau^{\prime}, \sigma^{\prime}\right)^2}\, \delta^{(2)}\left(z^{\prime}-z_k(s)\right),
\end{equation}
and we get
\begin{equation}
\begin{aligned}
\widehat{O}_{\text{defect}}[r ;\{\gamma_k\}] & =\frac{\pi \alpha'}{ \beta^2 \epsilon} \sum_{k=1} \int_0^{\ell_{\tau}} d s \int_0^{\ell_\tau} d \tau^{\prime} \int_0^{2 \pi} d \sigma^{\prime} \frac{\left|z_k(s)\right|}{r\left(\tau^{\prime}, \sigma^{\prime}\right)^2} \delta^{(2)}\left(z^{\prime}-z_k(s)\right) \\
& =\frac{\pi \alpha'}{\beta^2 \epsilon} \int_0^{\ell_{\tau}} d \tau^{\prime} \int_0^{2 \pi} d \sigma^{\prime} \frac{1}{r\left(\tau^{\prime}, \sigma^{\prime}\right)^2} \sum_{k=1}^{N_{\text{def}}} \int_0^1 d s\left|\dot{z}_k(s)\right| \delta^{(2)}\left(\left(\tau^{\prime}, \sigma^{\prime}\right)-z_k(s)\right).
\end{aligned}
\end{equation}
Integrating over the worldsheet gives
\begin{equation}\label{eq:O_defect}
\widehat{O}_{\text{defect}}[r ;\{\gamma_k\}]=\frac{\pi\alpha^{\prime}}{ \beta^2 \epsilon} \sum_{k=1}^n \int_0^1 d s \frac{\left|\dot{z}_k(s)\right|}{r\left(z_k(s)\right)^2},
\end{equation}
and \eqref{eq:Kcyl_line_configs} now takes the form of a sum of one-point functions of $\widehat{O}_{\text{defect}}[r ;\{\gamma_k\}]$
\begin{align}
K_{\mathrm{cyl}}\left(\beta, \ell_\tau, r_c\right)&=\sum_{\gamma_k(s)}\int_{\substack{r(0, \sigma) = r_c \\ r(\ell_\tau, \sigma) = r_c}} [\mathcal{D}r]\left[{\frac{1}{4\pi r\beta}}\right]\exp(- \frac{1}{4 \pi \alpha'} \int d^2z \, \partial r\partial r + O_{\text {defect }}[r ;\{\gamma_k\}]) \\
&= \sum_{\gamma_k(s)} \langle\widehat{O}_{\text{defect}}[r;\{\gamma_k\}]\rangle .
\end{align}



\subsection{Boundary conditions}\label{ssec:BC}
A single vertical defect sits at fixed $\sigma=\sigma_{k}$ and extends from $\tau=0$ to $\tau=\ell_\tau$.

Split $r(\tau,\sigma)$ into a zero mode $r_0$ and a small fluctuation $\eta(\tau,\sigma) \ll 1$ which averages to zero on the worldsheet
\begin{equation}\label{eq:zero_mode_split}
r(\tau,\sigma)\;=\;r_{0}\;+\;\sqrt{\alpha'}\eta(\tau,\sigma),
\qquad
r_{0}\;\hbox{constant over $\Sigma$},
\qquad
\int_{\Sigma}\eta=0. 
\end{equation}

Let $r_{c}$ be the fixed target space cutoff radius  
\begin{equation}\label{eq:radial_cutoff}
\,r(0,\sigma)=r(\ell_\tau,\sigma)=r_{c}\quad\forall\sigma.  
\end{equation}
Because $r_{0}$ is spatially constant, then
\begin{equation}\label{eq:eta_BC}
\eta(\tau{=}0,\sigma)=\eta(\tau{=}\ell_\tau,\sigma)=(r_0-r_c)/\sqrt{\alpha'}:=r_b. 
\end{equation}
Note that taking $r_b = 0$ means  $\,r(0,\sigma)=r(\ell_\tau,\sigma)=r_{0}\quad\forall\sigma\,$ and 
$\eta(\tau{=}0,\sigma)=\eta(\tau{=}\ell_\tau,\sigma)=0$.

\subsection{The measure factor}\label{ssec:measure_factor}
With the fluctuation mode $\eta(\tau, \sigma)$, the measure factor can be expressed as the following product over worldsheet points
\begin{equation}
\left[{\frac{1}{4\pi r\beta}}\right]= \prod_{\sigma, \tau} \frac{\sqrt{\alpha^{\prime}} d \eta(\tau, \sigma)}{4 \pi \beta\left[r_0+\sqrt{\alpha^{\prime}} \eta(\tau, \sigma)\right]} =\left[\prod_{\sigma, \tau} \frac{d \eta(\tau, \sigma)}{4 \pi \beta r_0}\right] \exp \left[-\sum_{\sigma, \tau} \ln \left(1+\frac{\sqrt{\alpha^{\prime}} \eta(\tau, \sigma)}{r_0}\right)\right] .
\end{equation}
Expanding $\ln \left[1+\frac{\sqrt{\alpha^{\prime}} \eta}{r_0}\right]$ in small $\eta(\tau, \sigma)$ 
\begin{equation}
\ln \left[1+\frac{\sqrt{\alpha^{\prime}} \eta}{r_0}\right]=\frac{\sqrt{\alpha^{\prime}} \eta}{r_0}-\frac{\alpha^{\prime} \eta^2}{2 r_0^2}+O\left(\eta^3\right),
\end{equation}
gives
\begin{equation}
\left[d \mu_r\right]=\left[\prod_{\sigma, \tau} \frac{\sqrt{\alpha'}d \eta}{4 \pi \beta r_0}\right] \exp \left[-\frac{\sqrt{\alpha^{\prime}}}{r_0} \sum_{\sigma, \tau} \eta+\frac{\alpha^{\prime}}{2 r_0^2} \sum_{\sigma, \tau} \eta^2+\ldots\right] .
\end{equation}

To enforce the boundary condition \eqref{eq:eta_BC} on $\eta(\tau, \sigma)$ for one line defect, we insert a Lagrange multiplier 
\begin{equation}\label{eq:Lagrange_Multiplier}
1=\int[\mathcal{D} \Lambda_0] \exp \left[i \int_0^{2\pi} d \sigma \Lambda_0(\sigma) (\eta\left(0, \sigma \right)-r_b)\right],\quad 1=\int[\mathcal{D} \Lambda_{\ell_{\tau}}] \exp \left[i \int_0^{2\pi} d \sigma \Lambda_T(\sigma)( \eta\left(\ell_\tau, \sigma\right) -r_b)\right].
\end{equation}

Writing the boundary condition as a product of delta functions 
\begin{equation}\label{eq:eta_BC_prod}
\prod_\sigma \delta\left(\eta\left(0, \sigma\right)-r_b\right), \quad \prod_\sigma \delta\left(\eta\left(\ell_\tau, \sigma\right)-r_b\right),
\end{equation}
the full measure reads
\begin{equation}\label{eq:measure_BC}
\left[\prod_{(\sigma, \tau) \in \Sigma} \frac{d \sqrt{\alpha'}\eta(\tau, \sigma)}{4 \pi \beta r_0}\right]\exp \left[\frac{\alpha^{\prime}}{2 r_0^2} \int_{\Sigma} \eta^2+O\left(\eta^3\right)\right] \prod_\sigma \delta\left(\eta\left(0, \sigma\right)-r_b\right)\, \prod_\sigma \delta\left(\eta\left(\ell_\tau, \sigma\right)-r_b\right) .
\end{equation}
Here, the \textit{bulk} points $(\sigma, \tau) \in \Sigma $ integrate freely and the linear term drops because $\int_{\Sigma } \eta=0$.


\subsection{The worldsheet action}
Now we further process the two terms in the effective worldsheet action 
\begin{equation}\label{eq:worldsheet_action}
I_{\text{ws}} =\frac{1}{4 \pi \alpha^{\prime}} \int d^2 z \, \partial r \partial r+\frac{\pi\alpha^{\prime}}{\beta^2 \epsilon} \sum_{k=1}^{N_{\text{def}}} \int_0^1 d s \frac{\left|\dot{z}_k(s)\right|}{r\left(z_k(s)\right)^2}.
\end{equation}

\medskip

\noindent Without loss of generality, we can assume that the line defect is vertical
\begin{equation}
z_k(s)=\left(\tau(s), \sigma_0\right), \quad \tau(s)=\ell_\tau\, s, \quad\left|\dot{z}_k(s)\right|=\ell_\tau,
\end{equation}
and thus
\begin{equation}
\int_0^1 d s\left|\dot{z}_k(s)\right|=\int_0^1 \ell_\tau \, d s=\ell_\tau.
\end{equation}

Therefore, $\widehat{O}_{\mathrm{defect}}$ \eqref{eq:O_defect} simplifies to
\begin{equation}
\widehat{O}_{\mathrm{defect}}=\frac{\pi \alpha^{\prime}}{\beta^2 \epsilon} \int_0^{\ell_\tau} d\tau\sum_{k=1}^{N_{\text{def}}} \frac{1}{r\left(\tau, \sigma_k\right)^2} .
\end{equation}
Splitting $r(\tau, \sigma)=r_0+\sqrt{\alpha^{\prime}} \eta(\tau, \sigma)$, we get
\be
\frac{1}{\left(r_0+\sqrt{\alpha^{\prime}} \eta\right)^2}=\frac{1}{r_0^2}\left(1-2 \frac{\sqrt{\alpha^{\prime}} \eta}{r_0}+3 \frac{\alpha^{\prime} \eta^2}{r_0^2}+O\left(\alpha^{\prime 3 / 2} \eta^3\right)\right) .
\ee
Then to $O\left(\eta^2\right)$, we have
\be
\frac{1}{r^2\left(\tau, \sigma_0\right)}=\frac{1}{r_0^2} -2 \frac{\sqrt{\alpha^{\prime}} \eta}{r_0^3}+3 \frac{\alpha^{\prime} \eta^2}{r_0^4}.
\ee

The worldsheet action now reads
\begin{equation}
I_{\text{ws}} =\frac{1}{4 \pi \alpha^{\prime}} \int d^2 z \, \partial r \partial r + \frac{\pi \alpha^{\prime}}{\beta^2 \epsilon} \sum_{k=1}^{N_{\text{def}}} \int_0^{\ell_{\tau}} d\tau \left(\frac{1}{r_0^2} -2 \frac{\sqrt{\alpha^{\prime}} \eta (\tau, \sigma_k)}{r_0^3}+3 \frac{\alpha^{\prime} \eta^2(\tau, \sigma_k)}{r_0^4} + O(\eta^3)\right).
\end{equation}

\subsection{The partition function}
In terms of $\eta(\tau,\sigma)$, the partition function for $N_{\text{def}}$ line defect insertions reads
\begin{equation}\label{eq:KK}
\begin{aligned}
K_{\mathrm{cyl}}\left(\beta, \ell_\tau, r_c\right)&=e^{-\alpha^{\prime} \ell_\tau N_{\text{def}} /\left(4 \beta^2 \epsilon r_0^2\right)}  \times \int\left[\prod_{(\sigma, \tau) \in \Sigma} \frac{\sqrt{\alpha'}d \eta}{4 \pi \beta r_0}\right]\left[\prod_{\sigma}d\Lambda\right]\\
&\exp\left(-i \int_0^{2\pi}\left( \Lambda_0(\sigma)(\eta\left(0, \sigma\right)-r_b) + \Lambda_0(\sigma)(\eta\left(\ell_\tau, \sigma\right)-r_b)\right) d \sigma\right)\\
&\exp \left[-\frac{1}{4 \pi}\int_{\Sigma }(\partial \eta)^2+\frac{\alpha^{\prime}}{2 r_0^2} \int_{\Sigma } \eta^2+\frac{\pi\alpha'}{\beta^2\epsilon r_0^2} \sum_{k} \int_{\gamma_k}\left(2\frac{\sqrt{\alpha^{\prime}} \eta}{r_0}-3\frac{\alpha'\eta^2}{r_0^2}\right)
+O\left(\eta^3\right)\right].
\end{aligned}
\end{equation}
Here, $\int_{\gamma_k}=\int_0^{\ell_\tau}d\tau|_{\sigma=\sigma_k}$ integrates over a vertical flux line in the $\tau$-direction. Notice that only $\sqrt{\alpha'}$ and $r_0$ are dimensionful; all other variables  are dimensionless.

\subsection{Fourier decomposition}
Equation \eqref{eq:eta_BC} tells us that there is no sudden jump of the field $r$ once we join the two ends of the worldsheet $\tau=0$ with $\tau=\ell_\tau$. This is the same as imposing periodic boundary condition for the interval $\tau\in[0,\ell_\tau]$. Therefore, for any configuration of $r$, a Fourier expansion gives
\begin{equation}\label{eq:fourier:eta}   \eta(\sigma,\tau)=\sum_{m,n}a_{m,n}\exp(i(n\sigma+\frac{2m\pi}{\ell_\tau}\tau)).
\end{equation}
Reality conditions give
\begin{equation}
    a_{-m,-n}=a^\dagger_{m,n}.
\end{equation}

We also Fourier expand the boundary Lagrange multiplier term
\begin{equation}
\int_0^{2\pi} \Lambda_0(\sigma)(\eta\left(0, \sigma\right)-r_b)=\sum_{m,n}(a_{m,n}-\delta_{n,0}r_b)\Lambda_n.
\end{equation}
Notice that equation \eqref{eq:fourier:eta} implies that only one set of Lagrange multipliers is needed. 

\medskip

In terms of modes, the bulk $O(\eta^2)$ term reads
\begin{equation}
  \exp\left( -\frac{1}{4 \pi} \int_{\Sigma }(\partial \eta)^2+\frac{\alpha^{\prime}}{2 r_c^2} \int_{\Sigma } \eta^2 \right) = \exp \left(A_\Sigma\sum_{m,n}|a_{m,n}|^2\left(-\frac{1}{4\pi}(n^2+(\frac{2\pi m}{\ell_\tau})^2)+\frac{\alpha'}{2r_c^2}\right)\right),
\end{equation}
where we realize that $A_\Sigma\left(\frac{1}{4\pi}(n^2+(\frac{2\pi m}{\ell_\tau})^2)-\frac{\alpha'}{2r_c^2}\right)$ is related to the mass$^2$ of the relevant excitation and the integral over $\eta^2$ gives the area of the worldsheet $A_{\Sigma}$ using the completeness relations of $A_{mn}$. Note that there would be tachyonic modes for small $n$ and $m$. When $r_0>\sqrt{2\pi \alpha'}$, all the $n\neq0$ modes are not tachyonic. Also, when $\ell_\tau<\sqrt{\frac{2\pi r_0^2}{\alpha'}}$, all $m\neq0$ modes are not tachyonic.

Line defects experience a strong mutual repulsive force, and tend to be equally distributed in the $\sigma$-direction. We can shift $\sigma$-coordinate and, without loss of generality, assume $\sigma_1=0$. Then the contribution from the linear term is the sum over the number of defect (flux) lines $N_{\text{def}}$
\begin{equation}
\sum_{k=1}^{N_{\text{def}}}\int_{\gamma_k}\eta(\sigma_k,\tau)d\tau=\ell_\tau\sum_{n,k}a_{0,n}\exp(in\sigma_k),
\end{equation}
vanishes unless $n\equiv 0 \mod N_{\text{def}}$, and 
\begin{equation}
\sum_{k=1}^{N_{\text{def}}}\int_{\gamma_k}\eta(\sigma_k,\tau)d\tau=\ell_\tau N_{\text{def}}\sum_n a_{0,nN_{\text{def}}}.
\end{equation}
Similarly, the second order term is
\begin{equation}
\sum_{k=1}^{N_{\text{def}}}\int_{\gamma_k}\eta^2(\sigma_k,\tau)d\tau=\ell_\tau N_{\text{def}}\sum_{m',n',n} a_{m',nN_{\text{def}}+n'}a_{-m',-n'}.
\end{equation}

Putting things together, the cylinder partition function in terms of the radial oscillator modes is the sum of \textit{all} possible configurations $\{\gamma_k\}$  of a \textit{gas} of $N_{\text{def}}$ line defect insertions 
\begin{equation}\label{eq:radial_prop}
\begin{aligned}
K_{\mathrm{cyl}}\left(\beta, \ell_\tau, r_c\right)= &\sum_{\{ N_{\text{def}}\}}  e^{-\alpha^{\prime} \ell_\tau N_{\text{def}} /\left(4 \beta^2 \epsilon r_0^2\right)}  \times \int\left[\prod_{m,n} \frac{\sqrt{\alpha'}d a_{m,n}}{4 \pi \beta r_0}\right]\left[\prod_{n}d\Lambda_n\right]\\
&\exp\left(-i\sum_{m,n}(a_{m,n}-\delta_{n,0}r_b)\Lambda_n\right)\\
&\exp \left[-A_\Sigma\sum_{m,n}|a_{m,n}|^2\left(\frac{1}{4\pi}(n^2+(\frac{2\pi m}{\ell_\tau})^2)-\frac{\alpha'}{2r_0^2}\right)\right]\\
&\exp\left[\frac{\pi \alpha' \ell_\tau N_{\text{def}}}{\beta^2\epsilon r_0^2} \left(2\frac{\sqrt{\alpha^{\prime}}}{r_0}\sum_n a_{0,nN_{\text{def}}}-3\frac{\alpha'}{r_0^2}\sum_{m',n',n} a_{m',nN_{\text{def}}+n'}a_{-m',-n'}\right) 
+O\left(\eta^3\right)\right].
\end{aligned}
\end{equation}
A nonzero contribution to the entropy will come entirely from the last term in \eqref{eq:radial_prop}. It is the only term that contains the expansion in $\eta(\sigma, \tau)$ of the inverse--square potential term, or equivalently, of the line defect operator $\eqref{eq:O_defect}$, and the only term with nontrivial dependence on $\beta$; it is a reflection of nontrivial (topological) winding sectors in $K_{\mathrm{cyl}}$. Without it, the string does not see the tip of the cone at $r_c$, and thus has zero contribution to the entropy. 

Let $T_{\mathrm{defect}}>0$ be the intrinsic tension (mass density) carried by each defect line
\begin{equation}\label{eq:Tline_def}
T_{\mathrm{defect}} \;=\,\frac{1}{\beta^2\,r_c^2}.
\end{equation}
Then the action density for the tension of $N_{\text{def}}$ line defects is
\begin{equation}\label{eq:I_tension}
I_{\mathrm{tension}} \;=\; N_{\text{def}} \,T_{\mathrm{defect}}\,\ell_\tau \,\frac{\alpha'}{\eps}.
\end{equation}
Note that \eqref{eq:I_tension} depends only on $N_{\text{def}}\,\ell_\tau$ and does not depend on $r(\tau, \sigma_k)$ of any line defect. When one sums over $N_{\text{def}}$, the path integral weight $\exp(-I_{\mathrm{tension}})$ \emph{penalizes} sectors with more line defects: each additional line contributes the same positive amount \eqref{eq:Tline_def} to the action. In this sense, the intrinsic defect line tension acts like a positive \textit{chemical potential} for the defect lines on the worldsheet, and thus the action for a larger number of defect insertions is exponentially suppressed, and it is therefore justifiable that we only compute the partition function in the zero-defect sector ($K_0$) and  the one-defect ($K_1$). With the line tension \eqref{eq:Tline_def}, the point $r_c$ can, in some sense, be interpreted as an \textit{off-shell} D0-brane \cite{Rychkov:2002ni}, and its image in target space is a D0-brane worldline. We will have more to say about this point in Section \ref{sec:discussion}.

\subsection{The Cylinder Amplitude}\label{ssec:Cylinder_Ampl}

The cylinder partition function $K_0$ for ${N_{\text{def}}} = 0$ is given by
\begin{equation}
\begin{aligned}
 K_0(\beta,\ell_\tau; r_b)=&\int\left[\prod_{m,n} \frac{\sqrt{\alpha'}d a_{m,n}}{4 \pi \beta r_0}\right]\left[\prod_{n}d\Lambda_n\right]\exp[ir_b\Lambda_0]\\&
\exp\left[-\sum_{m,n}(a_{m,n}+i\frac{\Lambda_{-n}}{M_{m,n}A_\Sigma})(a_{-m,-n}+i\frac{\Lambda_{n}}{M_{m,n}A_\Sigma})M_{m,n}A_\Sigma-\frac{\Lambda_n\Lambda_{-n}}{A_\Sigma M_{m,n}}\right]\\
=&\prod_{m,n}\frac{\sqrt{\alpha'/A_\Sigma}}{ M_{m,n}4\beta r_0}\int\left[\prod_n d\Lambda_n\right]\\&\exp\left[-\sum_n(\sum_m\frac{1}{M_{m,n}A_\Sigma})(\Lambda_n-\frac{ir_b\delta_{0,n}}{2\sum_{m\neq0}\frac{1}{A_\Sigma M_{m,n}}})(\Lambda_{-n}-\frac{ir_b\delta_{0,n}}{2\sum_{m\neq0}\frac{1}{A_\Sigma M_{m,n}}})\right]\exp(-\frac{r_b^2}{4\sum_{m\neq0}\frac{1}{A_\Sigma M_{m,0}}})\\
=&\prod_{m,n}\frac{\sqrt{\alpha'/A_\Sigma}}{M_{m,n}4\beta r_0}\prod_n\frac{\pi}{\sum_m \frac{1}{A_\Sigma M_{m,n}}}\exp(-\frac{r_b^2}{4\sum_{m\neq0}\frac{1}{A_\Sigma M_{m,0}}}).
    \end{aligned}
\end{equation}
Here $M_{m,n}:=\frac{1}{4\pi}(n^2+(\frac{2\pi m}{\ell_\tau})^2)-\frac{\alpha'}{2r_0^2}$ is the mass matrix and $m,n \in \mathbb{Z}$. On the lattice, $m,n$ will have a natural cutoff. However, in our calculation, $m,n$ takes all values in $\mathbb{Z}$, which amounts to taking the continuum limit, but keeping the lattice spacing $\eps$ as the UV cutoff.

\medskip

Details of calculating $K_0$ are given in Appendix \ref{app:K_0}. Below, we only present the details of the last step where we carry out the modular integral of $K_0(\beta,\ell_\tau; r_b)$ to obtain the cylinder amplitude and the entropy.

\medskip

Let $2\pi\alpha/r_0^2\coloneqq\mu^2$. Taking $r_b=0$, the modular integral can be directly evaluated for small $\mu$
\begin{equation}
K_0(\beta, \mu):=\int d\ell_\tau \,2\sqrt 2\pi \, \beta 
\,\left[
\ell_\tau\mu\sin\!\Bigl(\ell_\tau\, \mu\Bigr)\;
\frac{\eta\!\left(i\,\tfrac{\pi}{\ell_\tau}\right)^{3}}
{\vartheta_{1}\!\left(\ell_\tau \mu\,\middle|\, i\,\tfrac{\pi}{\ell_\tau}\right)}
\right]^{\!\tfrac12}.
\end{equation}
where $\eta(z)$ is the Dedekind eta function defined as $\eta(z)=q^{1/24}\prod_n(1-q^2)^n$. One has the classical identity for the Jacobi theta function $\vartheta_1$,
\begin{equation}
  \vartheta_1'(0|\tau) = 2\pi\,\eta(\tau)^3.
  \label{eq:theta1-prime-0}
\end{equation}
When $\ell_\tau \mu<<1$, $\frac{\eta(\tau)^3}{\vartheta_1(\ell_\tau\mu|\tau)}
  \;=\;
  \frac{1}{2\pi \ell_\tau\mu}\,\bigl(1+O((\ell_\tau\mu)^2)\bigr)$. Combined with the Taylor expansion for $\sin(\ell_\tau\mu)$, 
  \begin{equation}
      \ell_\tau\mu\sin\!\Bigl(\ell_\tau\, \mu\Bigr)\;
\frac{\eta\!\left(i\,\tfrac{\pi}{\ell_\tau}\right)^{3}}
{\vartheta_{1}\!\left(\ell_\tau \mu\,\middle|\, i\,\tfrac{\pi}{\ell_\tau}\right)}=\frac{\mu \ell_\tau}{2\pi}(1+O(\mu\ell_\tau)^2).
  \end{equation}
Therefore, the $\ell_\tau$ integral for small $\ell_\tau<R$, where R is a cutoff, evaluates as
\begin{equation}
2\sqrt{\pi\mu}\beta \int^R d \ell_\tau \sqrt{\ell_\tau}=\frac{4\sqrt{\pi\mu}}{3}\beta R^{3/2}.\label{eq:2.65}
\end{equation}
Above the cutoff $R$, we can use a modular transform of the Jacobi theta function as follows. Denote $\tau'=-\frac{1}{\tau}$ as the inversion of $\tau$, standard 
\begin{equation}
\frac{\eta(\tau)^3}{\vartheta_1(x|\tau)}=\frac{i}{(-i\tau)}\,e^{-\,i x^2/(\pi\tau)}\,\frac{\eta(\tau')^3}{\vartheta_1\!\bigl(x/\tau\,\big|\tau'\bigr)},
\end{equation}
then
\begin{equation}
  \biggl|
\frac{\eta\,\Bigl(\tfrac{i\pi}{\ell_\tau}\Bigr)^3}
{\vartheta_1\!\Bigl(\ell_\tau\mu\,\big|\,\tfrac{i\pi}{\ell_\tau}\Bigr)}\biggr|\;\le\;C\,\exp\,\Bigl(-\,\frac{\mu^2}{\pi^2}\ell_\tau^3\Bigr),\qquad \ell_\tau\to\infty,
\end{equation}
for some constants $C>0$. So the contribution of the $\ell_\tau$ integral when $\ell_\tau>R$ is bounded by 
\begin{equation}
2\sqrt{2}\pi\beta\int_0^\infty d\ell_\tau \, \bigg(\ell_\tau \,\mu|\,\sin(\ell_\tau \mu)|\bigg)^{1/2}\, C\,\exp\,\Bigl(-\,\frac{\mu^2}{2\pi^2}\ell_\tau^3\Bigr)<\frac{4C\pi^{5/2}\beta}{3}\mu^{-1/2},\label{eq:2.68}
\end{equation}
after using $|\sin (\ell_\tau \mu)|\leq1$. Combining \eqref{eq:2.65} and \eqref{eq:2.68}, with the selection $R=\mu^{-1}$,
\begin{equation}
    K_0(\beta,\mu)<\frac{4\sqrt{\pi}}{3}\beta\mu^{-1/2}(C\pi^2+1)=O(\mu^{-1/2}).
\end{equation}
In fact, if we plug in $R=\mu^{-1}$ as the lower bound in \eqref{eq:2.68}, we can obtain a tighter bound 
\begin{equation}
2\sqrt{2}\pi\beta\int_{\mu^{-1}}^\infty d\ell_\tau (\ell_\tau \mu)^{1/2} C\,\exp\!\Bigl(-\,\frac{\mu^2}{2\pi^2}\ell_\tau^3\Bigr)\approx \frac{2C\pi^2}{3}\exp(-\frac{1}{2\pi\mu^2}),
\end{equation}
so
\begin{equation}\label{eq:K_0}
K_0(\beta,\mu)=\frac{4\sqrt{\pi}}{3}\beta\mu^{-1/2}+O(\mu^{-3/2}).
\end{equation}

Since $K_0(\beta, \mu)$ is linear in $\beta$, the entropy vanishes as expected 
\begin{equation}\label{eq:K_0_entropy}
    S_0=(1-\beta\partial_\beta)|_{\beta=1}K_0(\beta,\mu)=0.
\end{equation}

\medskip

Observe that we do not constrain the fluctuation field $\eta(\sigma, \tau)$ to always stay above $r_b$; the string will inevitably cross below  $r=r_c$. However, there will be an \textit{equal} number of crossings from the left and right of $r_b$ such that the net number of crossings is zero. This is the main reason why the entropy vanishes in this zero-winding sector. In nontrivial topological sectors with nonzero winding, in contrast, the string will cross below $r_c$ an odd number of times, yielding nonzero entropy.



\medskip

Let us now consider $N_{\text{def}}=1$. Define $C:=\frac{3\alpha^{\prime 2}\ell_\tau}{4\beta^2\epsilon r_0^4}=\frac{3\mu^4}{16\pi^2\beta^2}\frac{\ell_\tau}{\epsilon}$ and the matrices $[L_m]_{n,n'}:=A_\Sigma M_{m,n}\delta_{n,n'}+C$, then
\begin{equation}\label{eq:K_1_alpha'}
\begin{aligned}
K_1(\beta, \ell_{\tau}, \eps)= &e^{-\alpha^{\prime} \ell_\tau /\left(4 \beta^2 \epsilon r_0^2\right)}  \times \int\left[\prod_{m,n} \frac{\sqrt{\alpha'}d a_{m,n}}{4 \pi \beta r_0}\right]\left[\prod_{n}d\Lambda_n\right]\\
&\exp\left(-i\sum_{m,n}a_{m,n}\Lambda_n\right)\exp(ir_b\Lambda_0)\\
&\exp \left[-\sum_{m,n,n'}a_{m,n}a_{-m,-n'}[L_m]_{n,n'}\right]\exp\left[\frac{2\sqrt{2\pi}}{3\mu}C\sum_n a_{0,n}
+O\left(\eta^3\right)\right].
\end{aligned}
\end{equation}

In Section \ref{sec:saddle_point}, we compute $K_1(\beta, \ell_{\tau},\eps)$ using saddle point approximation. In Part II of our work, we will implement the $\alpha'$-corrected $K_1$ \eqref{eq:K_1_alpha'}.

\section{The One-Line Defect Entropy From Saddle Point Analysis }\label{sec:saddle_point}

In this section, we restrict the 2D worldsheet theory to a single vertical defect line at fixed $\sigma=\sigma_k$. The resulting configuration describes a \emph{radial trajectory} $r(\tau)$ in a half-line, representing an \emph{open string} whose endpoints are fixed at $r=r_c$. We will use this 1D action to compute the off-shell classical (stationary) action $I_{\min}$, the one-defect partition function $K_1$, the cylinder amplitude, and the ultimately the entropy. We will explain how, by relating the UV and IR limits of the modulus integral, the resulting entropy can be rendered finite in a specific renormalization scheme. We will then argue in Section \ref{sec:constant_time_slice} that this open string lives on a constant Euclidean-time slice of the target space cone. 


\subsection{The 1D half-line action}
Consider a cylinder $\Sigma=[0,\ell_\tau]\times S^1_{\sigma}$ with coordinates $(\tau,\sigma)$, flat metric, and Dirichlet boundary conditions
\begin{equation}\label{eq:BC_2d}
r(0,\sigma)=r(\ell_\tau,\sigma)=r_c>0,\qquad r(\tau,\sigma+2\pi)=r(\tau,\sigma).
\end{equation}
The worldsheet action is
\begin{equation} \label{eq:Iws_2d}
I_{\mathrm{ws}}[r]=\frac{1}{4\pi\alpha'}\int_{0}^{\ell_\tau}\!\!\!d\tau\int_{0}^{2\pi}\!\!d\sigma\;\big[(\partial_\tau r)^2+(\partial_\sigma r)^2\big]
+\frac{\pi \alpha'}{\beta^2\,\epsilon}\int_{0}^{\ell_\tau}\!d\tau\;\frac{1}{r(\tau,\sigma_k)^2},
\end{equation}

Varying $r\mapsto r+\delta r$ with $\delta r|_{\tau=0,\ell_\tau}=0$ gives
\begin{equation}\label{eq:EOM_2d}
\delta I_{\mathrm{ws}}=\int_{0}^{\ell_\tau}\!\!\!d\tau\int_{0}^{L}\!\!d\sigma\;\delta r\;\Big\{
-\frac{1}{2\pi\alpha'}\Delta r(\tau,\sigma)
-\frac{2\pi\,\alpha'}{\beta^2\,\epsilon}\,\frac{\delta(\sigma-\sigma_k)}{r(\tau,\sigma_k)^3}
\Big\}=0,
\end{equation}
hence 
\begin{equation}\label{eq:PDE_2d}
\Delta r(\tau,\sigma)\;=\;-\lambda\;\frac{\delta(\sigma-\sigma_k)}{r(\tau,\sigma_k)^3}\,,\qquad
\lambda:=\frac{4\pi^{2}\,\alpha'{}^2}{\beta^2\,\epsilon}\,,
\qquad \Delta:=\partial_\tau^2+\partial_\sigma^2,
\end{equation}
with periodicity \eqref{eq:BC_2d}. Away from $\sigma=\sigma_k$, $r(\tau, \sigma)$ is a harmonic function. Integrating \eqref{eq:PDE_2d} across a thin strip $\sigma\in(\sigma_k-\eta,\sigma_k+\eta)$ and letting $\eta\to 0$ gives
\begin{equation}\label{eq:jump_sigma}
\partial_\sigma r(\tau,\sigma_k^{+})-\partial_\sigma r(\tau,\sigma_k^{-})
= -\,\lambda\,\frac{1}{r(\tau,\sigma_k)^3}.
\end{equation}
Thus, the stationary point is harmonic away from the line and has a controlled cusp (finite jump in normal derivative) at $\sigma=\sigma_k$.

\medskip

Let $G(\tau,\sigma;\tau',\sigma')$ solve
\begin{equation}\label{eq:Green_eq}
-\Delta G=\delta(\tau-\tau')\delta(\sigma-\sigma')\,,\qquad
G|_{\tau=0,\ell_\tau}=0,\quad G(\tau,\sigma+L;\tau',\sigma')=G(\tau,\sigma;\tau',\sigma').
\end{equation}
With $r_c$ a harmonic function and satisfying \eqref{eq:BC_2d}, the unique solution of \eqref{eq:PDE_2d} is
\begin{equation}\label{eq:integral_solution}
r(\tau,\sigma)=r_c-\lambda\int_{0}^{\ell_\tau}\!d\tau'\;G(\tau,\sigma;\tau',\sigma_k)\;\frac{1}{r(\tau',\sigma_k)^3}\,.
\end{equation}
Evaluating \eqref{eq:integral_solution} on the line defect gives a closed 1D nonlocal equation for the line profile $r(\tau):=r(\tau,\sigma_k)$
\begin{equation}\label{eq:line_integral_equation}
r(\tau)=r_c-\lambda\int_{0}^{\ell_\tau}\!d\tau'\;G(\tau,\tau')\;\frac{1}{r(\tau')^3}\,,
\qquad G(\tau,\tau'):=G(\tau,\sigma_k;\tau',\sigma_k).
\end{equation}
For a \emph{thin} defect, $G(\tau,\tau')$ is sharply peaked at $\tau=\tau'$ with total weight fixed by the regulated kernel
\begin{equation}\label{eq:self_kernel}
\int_{0}^{\ell_\tau}\!d\tau'\;G(\tau,\tau')\,f(\tau')\;\approx\;\mathcal{M}(\epsilon,L_\sigma)\,f(\tau)\,,
\qquad \mathcal{M}(\epsilon,L):=\int_{-\infty}^{+\infty}\!ds\;G(\tau,\tau+s),
\end{equation}
where $\mathcal{M}(\epsilon,2\pi)$ encodes the short distance regularization around $\sigma_k$. Then \eqref{eq:line_integral_equation} reduces to the ODE
\begin{equation}\label{eq:local_ODE}
\ddot r(\tau)=-\kappa\,\frac{1}{r(\tau)^3}\,,
\qquad 
\kappa:=\frac{\lambda}{\mathcal{N}}\,,\quad \mathcal{N}:=\frac{1}{2\pi\alpha'}\;\frac{1}{\mathcal{M}(\epsilon,2\pi)}\,.
\end{equation}

\subsection{The stationary 1D (open string) action and partition function}\label{ssec:1D_action_saddle}
Consider the action for a single vertical defect line localized at a fixed worldsheet spatial coordinate $\sigma=\sigma_k$. The radial degree of freedom $r(\tau):=r(\tau,\sigma_k)$ is integrated over $\sigma$ across a thin tube of width $\sim \epsilon$ centered at $\sigma_k$ 
\begin{equation}\label{eq:Seff}
I_{\mathrm{1D}}[r]
=\int_{0}^{\ell_\tau}\!\Big[\,\frac{\epsilon}{4\pi\alpha'}\,\dot r(\tau)^2 \;+\; \frac{\pi\,\alpha'}{\beta^2\,\epsilon}\,\frac{1}{r(\tau)^2}\,\Big]\,d\tau,
\qquad r(0)=r(\ell_\tau)=r_c,
\end{equation}
The $V_{\text {eff }}(r) \propto 1 / r^2>0$ is a inverse-square repulsive potential, which prevents the radial trajectory $r(\tau)$ from venturing into the small-$r$ region.

The Lagrangian density in \eqref{eq:Seff} is $L(r,\dot r)=\dfrac{\epsilon}{4\pi\alpha'}\,\dot r^{\,2}+\dfrac{\pi\,\alpha'}{\beta^2\epsilon}\,\dfrac{1}{r^2}$. Its Euler–Lagrange equation is
\begin{equation}\label{eq:EOM}
\frac{d}{d\tau}\Big(\frac{\partial L}{\partial \dot r}\Big)-\frac{\partial L}{\partial r}
= \frac{d}{d\tau}\!\Big(2\frac{\epsilon}{4\pi\alpha'}\,\dot r\Big) - \Big(-2\frac{\pi\,\alpha'}{\beta^2\epsilon}\,\frac{1}{r^3}\Big)=0
\;\;\Longrightarrow\;\;
\ddot r(\tau)=-\frac{4\pi^{2}\,\alpha'{}^2}{\beta^2\,\epsilon^2}\,\frac{1}{r(\tau)^3}\,<\,0\,.
\end{equation}
The right-hand side is negative for every $\tau$, so $r(\tau)$ bends downward everywhere. With both endpoints fixed at $r_c$, a downward-bending solution must lie at or above $r_c$ everywhere along the path.

Because $L$ has no explicit $\tau$-dependence, the total energy density
\begin{equation}\label{eq:EnergyConst}
\mathcal E \;:=\; \dot r\,\frac{\partial L}{\partial \dot r}-L
\;=\; 2\frac{\epsilon}{4\pi\alpha'}\,\dot r^{\,2} - \Big(\frac{\epsilon}{4\pi\alpha'}\,\dot r^{\,2} + \frac{\pi\,\alpha'}{\beta^2\epsilon}\,\frac{1}{r^2}\Big)
\;=\; \frac{\epsilon}{4\pi\alpha'}\,\dot r^{\,2} - \frac{\pi\,\alpha'}{\beta^2\epsilon}\,\frac{1}{r^2},
\end{equation}
is constant along the solution. Let $R$ be the (unique) maximum of $r(\tau)$ at $\tau=\ell_\tau/2$ 
\begin{equation}\label{eq:Rdef}
R := r\!\left(\frac{\ell_\tau}{2}\right)\quad \text{so that}\quad \dot r\!\left(\frac{\ell_\tau}{2}\right)=0,\quad R>r_c.
\end{equation}
Evaluating \eqref{eq:EnergyConst} at $\tau=\ell_\tau/2$ gives
\begin{equation}\label{eq:EnergyValue}
\mathcal E \;=\; -\,\frac{\pi\,\alpha'}{\beta^2\epsilon}\,\frac{1}{R^2}\,.
\end{equation}
Subtracting \eqref{eq:EnergyValue} from \eqref{eq:EnergyConst} yields the \emph{first integral}
\begin{equation}\label{eq:FirstIntegral}
\dot r(\tau)^2 \;=\; \frac{4\pi^{2}\,\alpha'{}^2}{\beta^2\,\epsilon^2}\Big(\frac{1}{r(\tau)^2}-\frac{1}{R^2}\Big)
\;=\; \frac{4\pi^{2}\,\alpha'{}^2}{\beta^2\,\epsilon^2}\,\frac{R^2-r(\tau)^2}{R^2\,r(\tau)^2}\ .
\end{equation}
Eq. \eqref{eq:FirstIntegral} reduces the second-order ODE to a first-order relation involving the single constant $R$.

On the half-interval $\tau\in[0,\ell_\tau/2]$, the radius increases from $r_c$ to $R$, so we take the positive square root of \eqref{eq:FirstIntegral}
\begin{equation}\label{eq:drdtPositive}
\dot r(\tau) \;=\; \frac{2\pi\,\alpha'}{\beta\,\epsilon}\,\frac{\sqrt{R^2-r(\tau)^2}}{R\,r(\tau)}\,,\qquad 0\le \tau\le \frac{\ell_\tau}{2}.
\end{equation}
Separating variables and integrating from $\tau=0$ (where $r=r_c$) to $\tau=\ell_\tau/2$ (where $r=R$), we get
\begin{equation}\label{eq:EndpointIntegral}
\int_{r_c}^{R} \frac{R\,r}{\sqrt{R^2-r^2}}\,dr \;=\; \frac{2\pi\,\alpha'}{\beta\,\epsilon}\int_{0}^{\ell_\tau/2}\! d\tau\;=\;\frac{\pi\,\alpha'}{\beta\,\epsilon}\,\ell_\tau\,.
\end{equation}
The elementary integral evaluates to (substituting $r=R\sin\phi$)
\begin{equation}\label{eq:ElemInt}
\int_{r_c}^{R} \frac{r}{\sqrt{R^2-r^2}}\,dr \;=\; \sqrt{R^2-r_c^2}\,,
\end{equation}
so \eqref{eq:EndpointIntegral} becomes
\begin{equation}\label{eq:Rcondition}
R\,\sqrt{R^2-r_c^2}\;=\;\frac{\pi\,\alpha'}{\beta\,\epsilon}\,\ell_\tau\ .
\end{equation}
This determines $R$ uniquely from $(\ell_\tau,\beta,r_c)$.\footnote{Notice that $R=r_c$ is allowed only if $\ell_\tau=0$ (a zero-height cylinder).} To solve for $R$, set $y=R^2$ in $y(y-r_c^2)=\dfrac{4\pi^{2}\,\alpha'{}^2}{\beta^2\epsilon^2}\,\ell_\tau^2$ to find
\begin{equation}\label{eq:Rexplicit}
R^2 \;=\; \frac{1}{2}\left(r_c^2 + \sqrt{\,r_c^4 + \dfrac{4\pi^{2}\,\alpha'{}^2}{\beta^2\epsilon^2}\,\ell_\tau^2\,}\ \right),\qquad R>r_c\,. 
\end{equation}
An explicit closed form for the entire profile follows by integrating \eqref{eq:FirstIntegral} once more. Setting $u(\tau):=r(\tau)^2$ one finds
\begin{equation}\label{eq:Profile}
r(\tau) \;=\; \sqrt{\,R^2 - \Big(\frac{2\pi\,\alpha'}{\beta\,\epsilon}\,\frac{\tau-\ell_\tau/2}{R}\Big)^2\,}\,,\qquad 0\le \tau\le \ell_\tau\,,
\end{equation}
and it is easy to check that $r(0)=r(\ell_\tau)=r_c$ holds if and only if \eqref{eq:Rcondition} is satisfied. So, the saddle path starts at $r_c$, rises smoothly to $R$ at $\tau=\ell_\tau/2$, then folds back to $r_c$. Because $\ddot r<0$, the interior radius never goes below $r_c$ at the saddle.

Inserting \eqref{eq:FirstIntegral} into \eqref{eq:Seff} to eliminate $\dot r$ gives
\begin{equation}\label{eq:IntegrandSimplify}
\frac{\epsilon}{4\pi\alpha'}\,\dot r^{\,2} + \frac{\pi\,\alpha'}{\beta^2\epsilon}\,\frac{1}{r^2}
= \frac{\pi\,\alpha'}{\beta^2\epsilon}\,\frac{2}{r^2} - \frac{\pi\,\alpha'}{\beta^2\epsilon}\,\frac{1}{R^2}\,.
\end{equation}
Hence
\begin{equation}\label{eq:SminSplit}
I_{\min}
= \int_{0}^{\ell_\tau}\!\Big[\frac{\pi\,\alpha'}{\beta^2\epsilon}\,\frac{2}{r(\tau)^2} - \frac{\pi\,\alpha'}{\beta^2\epsilon}\,\frac{1}{R^2}\Big]\,d\tau
= 2\int_{0}^{\ell_\tau/2}\!\Big[\frac{\pi\,\alpha'}{\beta^2\epsilon}\,\frac{2}{r(\tau)^2} - \frac{\pi\,\alpha'}{\beta^2\epsilon}\,\frac{1}{R^2}\Big]\,d\tau.
\end{equation}
Using \eqref{eq:drdtPositive} to change variables
\begin{equation}\label{eq:ChangeVar}
d\tau \;=\; \frac{\beta\,\epsilon}{2\pi\,\alpha'}\,\frac{R\,r}{\sqrt{R^2-r^2}}\,dr\,.
\end{equation}
Substituting \eqref{eq:ChangeVar} into \eqref{eq:SminSplit} yields
\begin{equation}\label{eq:SminIntegral}
I_{\min}(R)
= \frac{R}{\beta}\int_{r_c}^{R}\!\frac{1}{\sqrt{R^2-r^2}}
\left(\frac{2}{r} - \frac{r}{R^2}\right)dr\,.
\end{equation}
The two elementary integrals are (with $r=R\sin\phi$)
\begin{equation}\label{eq:TwoInts}
\int_{r_c}^{R}\!\frac{dr}{r\sqrt{R^2-r^2}}
= \frac{1}{R}\ln\!\frac{R+\sqrt{R^2-r_c^2}}{r_c},
\qquad
\int_{r_c}^{R}\!\frac{r\,dr}{\sqrt{R^2-r^2}}
= \sqrt{R^2-r_c^2}.
\end{equation}
Plugging \eqref{eq:TwoInts} into \eqref{eq:SminIntegral} yields the minimal (stationary) action
\begin{equation}\label{eq:SminFinal}
I_{\min}(R)
= \frac{1}{\beta}\,\ln\!\frac{R+\sqrt{R^2-r_c^2}}{r_c}
\;-\; \frac{1}{2\beta}\,\frac{\sqrt{R^2-r_c^2}}{R}\,,
\qquad R\ \text{fixed by}\ \eqref{eq:Rcondition}. 
\end{equation}
The semiclassical contribution $K_1$ to the partition function is therefore
\begin{equation}\label{eq:K_1_classical}
K_1 (\beta, R, \eps) \coloneqq \exp(-I_{\min})
= \exp\!\left[
-\left(
\frac{1}{\beta}\,\ln\!\frac{R+\sqrt{R^2-r_c^2}}{r_c}
- \frac{1}{2\beta}\,\frac{\sqrt{R^2-r_c^2}}{R}
\right)\right]. 
\end{equation}

To calculate the semiclassical contribution to the cylinder amplitude, we integrate \eqref{eq:K_1_classical}  over $\ell_\tau$
\begin{equation}\label{eq:Zmod_def}
\mathcal{Z}_1(\beta, \eps) \coloneqq \int_{0}^{\infty} d\ell_\tau\; K_1 (\beta, \ell_{\tau}, \eps)\,,
\end{equation}
with $I_{\min}$ given by \eqref{eq:SminFinal}.

\medskip

We first express $\mathcal{Z}$ \eqref{eq:Zmod_def} in terms of $\ell_\tau$ using  \eqref{eq:Rcondition}
\begin{equation}\label{eq:ell_of_R}
R\,\sqrt{R^2-r_c^2}\;=\;\frac{\pi\,\alpha'}{\beta\,\epsilon}\,\ell_\tau
\quad\Longleftrightarrow\quad
\ell_\tau(R)\;=\;\frac{\beta\,\epsilon}{\pi\,\alpha'}\,R\sqrt{R^2-r_c^2}\,,\quad R\in[r_c,\infty)\,.
\end{equation}
Differentiating \eqref{eq:ell_of_R} gives the Jacobian
\begin{equation}\label{eq:Jac_R}
\frac{d\ell_\tau}{dR}
=\frac{\beta\,\epsilon}{\pi\,\alpha'}\,
\frac{2R^2-r_c^2}{\sqrt{R^2-r_c^2}}\,.
\end{equation}
Hence \eqref{eq:Zmod_def} becomes the $R$–integral
\begin{equation}\label{eq:Zmod_R}
\mathcal{Z}_1(\beta, R, \eps)
=\frac{\beta\,\epsilon}{\pi\,\alpha'}\,
\int_{R=r_c}^{\infty}\! dR\;
\frac{2R^2-r_c^2}{\sqrt{R^2-r_c^2}}\;
\exp\!\Big(-I_{\min}(R)\Big)\,,
\end{equation}
where $I_{\min}(R)$ is the minimal action \eqref{eq:SminFinal} expressed as a function of $R$
\begin{equation}\label{eq:Imin_of_R}
I_{\min}(R)
= \frac{1}{\beta}\,\ln\!\frac{R+\sqrt{R^2-r_c^2}}{r_c}
\;-\; \frac{1}{2\beta}\,\frac{\sqrt{R^2-r_c^2}}{R}\,.
\end{equation}

\medskip

Let
\begin{equation}\label{eq:y_def}
y:=\frac{R}{r_c}\ \ (\ge 1),\qquad \sqrt{R^2-r_c^2}=r_c\sqrt{y^2-1}.
\end{equation}
Then $dR=r_c\,dy$, and \eqref{eq:Zmod_R} becomes
\begin{equation}\label{eq:Zmod_y}
\mathcal{Z}_1 (\beta, y, \eps)
=\frac{\beta\,\epsilon\,r_c^2}{\pi\,\alpha'}\,
\int_{1}^{\infty}\! dy\;
\frac{2y^2-1}{\sqrt{y^2-1}}\;
\exp\!\Big(-I_{\min}(y)\Big)\,,
\end{equation}
with
\begin{equation}\label{eq:Imin_y}
I_{\min}(y)
=\frac{1}{\beta}\left[
\ln\!\big(y+\sqrt{y^2-1}\big)
-\frac{1}{2}\,\frac{\sqrt{y^2-1}}{y}\right].
\end{equation}
We can linearize the $I_{\text{min}}(y)$ by the following change of variables to the hyperbolic coordinate $u$\footnote{The hyperbolic coordinate $u$ can be interpreted as the rapidity of the point $y=R/r_c$.}. Let
\begin{equation}\label{eq:u_def}
y=\cosh u,\qquad u\in[0,\infty)\,.
\end{equation}
Then
\begin{equation}\label{eq:hyperbolic_id}
\sqrt{y^2-1}=\sinh u,\quad
\ln\!\big(y+\sqrt{y^2-1}\big)=\ln(\cosh u+\sinh u)=u,\quad
\frac{\sqrt{y^2-1}}{y}=\tanh u.
\end{equation}
Also
\begin{equation}\label{eq:ell_u}
y\sqrt{y^2-1}=\cosh u\,\sinh u=\tfrac{1}{2}\sinh 2u,
\qquad
\frac{d\ell_\tau}{du}=\frac{\beta\,\epsilon\,r_c^2}{\pi\,\alpha'}\,\cosh 2u
\quad\text{(from \eqref{eq:ell_of_R})}.
\end{equation}
With \eqref{eq:hyperbolic_id}, the minimal action \eqref{eq:Imin_y} linearizes
\begin{equation}\label{eq:Imin_u}
I_{\min}(u)=\frac{1}{\beta}\left[u-\frac{1}{2}\tanh u\right].
\;
\end{equation}
Therefore, \eqref{eq:Zmod_def} takes the form
\begin{equation}\label{eq:Zmod_u}
\mathcal{Z}_1 (\beta, u, \eps)
=\frac{\beta\,\epsilon\,r_c^2}{\pi\,\alpha'}\,
\int_{0}^{\infty}\! du\;\cosh(2u)\;
\exp\!\left\{-\,\frac{1}{\beta}\left[u-\frac{1}{2}\tanh u\right]\right\}.
\end{equation}
\eqref{eq:Zmod_u} shows that the amplitude is the sum over radial excursions of different lengths in target space: the string leaves $r=r_c$, reaches the radius $R=r_c\cosh u$, and returns. Large $u$ means $R\gg r_c$ (a very long outward excursion), while small $u$ means $R\approx r_c$ (a short hop). The Jacobian factor $\cosh(2u)$ counts the density of such configurations as a function of $u$, tending to favor longer excursions, whereas the exponential $e^{-I_{\min}(u)}$ suppresses them with a cost that grows essentially linearly in $u$. The infrared cutoff (e.g. $\ell_\tau\le \Lambda$ or, equivalently, $u\le u_{\max}$ by \eqref{eq:ell_u}) limits how far out radially the trajectory may wander before it turns back; the ultraviolet endpoint $u\to 0$ corresponds to $\ell_\tau\to 0$, i.e.\ a vanishingly short excursion that contributes trivially\footnote{The UV limit ($\ell_\tau\to 0$) is rather trivial since as we shrink the cylinder to zero length, the corresponding line defect degenerates into a point (a puncture) on a large circle in the worldsheet, and thus, the contribution to the partition function vanishes $\mathcal{Z}_{\mathrm{UV}}\;\sim\;\int_{0}^{\ell_0}\!d\ell_\tau\,e^{-\mu\,\ell_\tau}
=\frac{1-e^{-\mu\,\ell_0}}{\mu}\xrightarrow{\ \ell_0\to 0\ } 0$. In this limit, one should account for the worldsheet action dependence on $\sigma$.}. The observed behavior of $\mathcal{Z}_1$ is thus the outcome of a competition between measure growth $\propto\cosh(2u)$ and the Boltzmann factor $\propto e^{-I_{\min}(u)}$; the cone angle $\beta$ is what ultimately determines the overall behavior of $\mathcal{Z}_1$ in the IR.

\medskip

It may be insightful to contrast this behavior of $\mathcal{Z}_1$ with the zero-defect case $K_0$ in \eqref{eq:K_0} in terms of the saddle point approximation. Recall that $K_0$, as we explained in Section \ref{sec:Calc} has zero entropy. Without a line defect potential $1/r^2$, the equation of motion is that of a free particle 
\begin{equation}
\label{eq:EOM_zero_defect}
\ddot r(\tau)=0,
\end{equation}
so the classical solution  is just $r_c$
\begin{equation}
\label{eq:rcl_zero_defect}
r_{\mathrm{cl}}^{(0)}(\tau)\equiv r_c,
\end{equation}
and the mimimal action vanishes
\begin{equation}
\label{eq:Imin_zero_defect}
I_{\min}^{(0)}(\ell_\tau)\equiv 0.
\end{equation}
Physically, this is a string \textit{sitting straight} at $r=r_c$: there is no force pulling it away from $r_c$, so the classical path does not explore the bulk.

At the semiclassical level, the partition function is therefore
\begin{equation}
\label{eq:K0_def}
K_0(\beta,\ell_\tau)
= \mathcal \,e^{-I_{\min}^{(0)}(\ell_\tau)}
= 1,
\end{equation}
and has no dependence on $\beta$ or $\ell_\tau$. The corresponding cylinder amplitude after integrating over the modulus is
\begin{equation}
\label{eq:Z0_def}
\mathcal{Z}_0(\beta;\Lambda)
:=\int_0^\Lambda d\ell_\tau\;K_0(\beta,\ell_\tau)
=\Lambda,
\end{equation}
where $\Lambda$ is an IR cutoff. This contribution is independent of the cone angle and is simply the infinite volume of the moduli space. In practice, one subtracts this trivial background piece (or equivalently defines the tip--localized entropy \emph{relative} to the $\beta$--independent sector), such that the trivial topological sector has a vanishing contribution to the entropy (after renormalization).

\medskip

Next we calculate the $\mathcal{Z}_1$ amplitude.

\subsection{The $\mathcal{Z}_1$ amplitude}\label{ssec:SC_amplitude}

In the IR (long cylinder) limit,  $\ell_\tau\to\infty$, ($u\to\infty$), $\tanh u=1-2e^{-2u}+\cdots$ and $\cosh 2u=\tfrac12 e^{2u}[1+\mathcal{O}(e^{-4u})]$, so
\begin{equation}\label{eq:Imin_large_u_new}
I_{\min}(u)=\frac{1}{\beta}\left[u-\frac{1}{2}+\mathcal{O}(e^{-2u})\right],
\qquad
\cosh(2u)\,e^{-I_{\min}(u)}\sim \frac{1}{2}\,e^{1/(2\beta)}\,e^{\,(2-\frac{1}{\beta})u}.
\end{equation}
Using
\begin{equation}\label{eq:ell_large_u_new}
\ell_\tau(u)=\frac{\beta\,\epsilon\,r_c^{2}}{2\pi\,\alpha'}\,\sinh(2u)
\sim \frac{\beta\,\epsilon\,r_c^{2}}{4\pi\,\alpha'}\,e^{2u}
\quad\Rightarrow\quad
u\sim \frac{1}{2}\ln\!\left(\frac{4\pi\,\alpha'}{\beta\,\epsilon\,r_c^{2}}\,\ell_\tau\right),
\end{equation}
we obtain the IR limit of $K_1(\beta, \ell_{\tau}, \eps)$
\begin{equation}\label{eq:eIm_IR_new}
K_1(\beta, \ell_{\tau}, \eps) = e^{-I_{\min}(\ell_\tau)}\ \sim\ 
C(\beta,\epsilon)\ \ell_\tau^{-\,\frac{1}{2\beta}},
\qquad
C(\beta,\eps):=e^{\frac{1}{2\beta}}\left(\frac{\beta\,\epsilon\,r_c^{2}}{4\pi\,\alpha'}\right)^{\!\frac{1}{2\beta}}. 
\end{equation}

Let $\Delta(\beta) = \frac{1}{2\beta}$ be the scaling dimension of the defect line operator. Then $\mathcal{Z}_{\text{IR}}(\beta,\Lambda, \eps)$ has the following  divergence pattern that depends on the given range of $\beta$
\begin{equation}\label{eq:IR_integral_behavior_new}
\mathcal{Z}_1(\beta,\Lambda, \eps) = \int^{\Lambda}\!d\ell_\tau\, K_1(\beta, \ell_{\tau}, \eps)
\;\sim\;
\begin{cases}
\dfrac{C(\beta,\eps)}{1-\frac{1}{2\beta}}\ \Lambda^{\,1-\Delta(\beta)}, & \beta>\dfrac{1}{2}\quad(\text{power law divergence}),\\[8pt]
C(\beta,\eps)\,\ln\Lambda, & \beta=\dfrac{1}{2}\quad(\text{log divergence}),\\[8pt]
\dfrac{C(\beta,\eps)}{\Delta(\beta)-1}, & \beta<\dfrac{1}{2}\quad(\text{finite in $\Lambda$}).
\end{cases}
\end{equation}

Let $\frac{\alpha'}{4} M^2_{\text{eff}} = (\Delta(\beta) -1)$ the effective mass of a winding mode\footnote{We believe it can't be the bulk tachyon since $M_{\text {eff}}$ depends in the cone angle $\beta$.} localized to the boundary (line defect). Then 
\begin{itemize}
    \item $M_{\text {eff }}<0 \,\, (\Delta(\beta) < 1)$ indicates the presence of a \textit{tachyonic} mode in the spectrum, and hence the IR power-law divergence; the line defect operator is relevant. In this regime, one may say that a Hagedorn transition occurs at very low local temperature (since $T_{\text{loc}}(r) = \frac{1}{\beta r}$). The critical behavior in Rindler space has been studied by analyzing the string one-loop free energy (the thermal scalar action) \cite{Mertens-RandomWalk:2013}. For bosonic strings, it was observed that there would be a divergence in thermodynamical quantities for any value of $\beta$ if one includes $\alpha'$ corrections in the free energy, which implies the Hagedorn temperature is effectively zero. This seems to be consistent, in sime sense, with our finding in this regime. However, for bosonic strings, the analysis is more subtle and we don't quite understand how or if this dependence on $\beta$ is related to a critical behavior in Rinder space.
    \item $M_{\text {eff }} =0 \,\, (\Delta(\beta) =1)$ indicates the the presence of a \textit{massless} mode, thus the log divergence; the line operator is marginal.
    \item $M_{\text {eff }} > 0 \,\, (\Delta(\beta) > 1)$ indicates the presence of a \textit{massive} mode, and hence the convergence (decay) of  $\mathcal{Z}_{\text{IR}}(\beta,\Lambda, \eps)$; the defect operator is irrelevant. 
\end{itemize}


\medskip

For $\beta=1$ (Rindler space), we find
\begin{equation}\label{eq:IR_beta1}
\mathcal{Z}_1(1, \Lambda, \eps)\sim 2 \,e^{\frac{1}{2}}\left(\frac{\epsilon\,r_c^{2}}{4\pi\,\alpha'}\right)^{\frac{1}{2}}\ \Lambda^{1/2}\quad\text{(power divergence)}.
\end{equation}
For $\beta<1$ (conical deficit), $\mathcal{Z}_{\text{IR}}(\beta,\Lambda,\eps)$ diverges according to a power-law. Similarly, a conical surplus ($\beta > 1$) also diverges but slightly faster. The amplitude converges only in the range $ 0 <\beta < 1/2$. We think that this makes sense because the cone is approaching a half-line in this range. The logarithmic divergence at $\beta = 1/2$ seems to suggest there is fixed point (CFT) for $\mathbb{C}/Z_2$ orbifold ($N=2$). In nine spatial dimensions, it has been used to compute the one-loop correction to the conical entropy in string theory in half-space \cite{He:2014gva}. It would be interesting to understand this renormalization group flow picture to that studied in \cite{RamosRicciFlow:2017} and further commneted on in  \cite{ahmadain2022off}.
 
\medskip

It may be insightful to express $Z_1(\beta ; \Lambda, \eps)$ in terms of the hyperbolic coordinate $u \rightarrow \infty$ \eqref{eq:ell_large_u_new}
\begin{equation}
\mathcal{Z}_1(\beta ; \Lambda, \eps) \sim \frac{1}{2}\, e^{\Delta(\beta)}\int^{u_{IR}}\,\,d u\, e^{-u \, (2 \Delta(\beta) -2}) \,d u = \frac{1}{2}\, e^{\Delta(\beta)}\int^{u_{IR}} \, e^{-u \, L_0(\beta)} = \frac{1}{L_0(\beta)},
\end{equation}
which, modulo the factor $e^{\Delta(\beta)}$, takes the form of an open string propagator where $u$ is the gluing parameter of the worldsheet and $L_0(\beta)$ is the Hamiltonian \cite{WittenNoteRiemannSurfaces2012,Witten:SuperStringTheoryRevisted2019}.\footnote{We are omitting the $b_0$ mode in the propagator.}

\medskip

We are now to calculate the entropy by varying the cone angle $\beta$. Starting from the IR–regulated amplitude \eqref{eq:IR_integral_behavior_new}
\begin{equation}
\mathcal{Z}_1(\beta,\Lambda,\eps) = \frac{C(\beta,\eps)}{1-\frac{1}{2\beta}}\ \Lambda^{\,1-\frac{1}{2\beta}},
\end{equation}
and acting by $(1-\beta\,\partial_\beta)$ on $\mathcal{Z}(\beta,\Lambda,\eps)$ gives the entropy contribution of one defect
\begin{equation}\label{eq:SIR_beta1_Lambda}
S_1(r_c,\Lambda, \epsilon)
=(1-\beta\,\partial_\beta) \,\mathcal{Z}_1(\beta,\Lambda,\eps)\Big|_{\beta = 1} = 2e^{\frac12}\sqrt{\frac{\epsilon r_c^2}{4\pi\alpha'}}
\Lambda^{1/2}
\Bigg[2+\frac{1}{2}\ln\Big(\frac{\epsilon r_c^2}{4\pi\alpha'}\Big)-\frac{1}{2}\ln\Lambda\Bigg].
\end{equation}
Now we make the important observation that divergence in \eqref{eq:SIR_beta1_Lambda} splits up as the product of $\sqrt{\eps} \sqrt{\Lambda}$.  While it may not to natural to correlate the UV and IR limits of the moduli integral unless there there is \textit{good} reason to do so\footnote{We thank Edward Witten for pointing this out for us.}, we believe in this case, we do a very obvious one: to make the entropy finite! If we choose 
\begin{equation}\label{eq:RG_scheme_UV_IR}
\Lambda = \frac{\kappa^2}{\eps},
\end{equation}
where $\kappa$ is some unfixed constant. Then the entropy in the one-defect sector is
\begin{equation}\label{eq:SIR_beta1_eps_only}
S_1(r_c,\epsilon)
=\kappa\frac{e^{\frac12}r_c}{\sqrt{\pi\alpha'}}
\Bigg[2+\ln\epsilon+\frac{1}{2}\ln\Big(\frac{r_c^2}{4\pi\alpha'}\Big)\Bigg].
\end{equation}

Acting with Tseytlin's \textbf{T1} prescription $\FTP$ on \eqref{eq:SIR_beta1_eps_only} to remove the remaining logarithmic divergence and extract the coefficient\footnote{The coefficient of $\ln \eps$ is normally the beta-function of some worldsheet operator, which in our case is the line defect, although this is not explicit in our calculation.} and choosing $\kappa =1$ (after absorbing the $e^{\frac12}$ factor), we obtain the entropy contribution of one line defect (an open string) localized at the point $r_c$\footnote{At first sight, it appears that $S = \mathcal{Z}(1)$ \eqref{eq:IR_beta1} after applying the RG scheme but this is not true for $\beta \neq 1$ $S_1(\beta)=\mathcal{Z}_1(\beta,\Lambda,\eps)\left(\frac{2\beta}{2\beta-1}+\frac{1}{2\beta}\ln(\frac{\epsilon r_c^2}{4\pi\alpha'\Lambda})\right)
$ and thus, the entropy and the cylinder amplitude do not agree away from $\beta=1$If we now make $S(\beta)$ finite, a natural choice which corresponds to \eqref{eq:RG_scheme_UV_IR} at $\beta =1$, is $\epsilon^{1/2\beta}\Lambda^{1-1/2\beta}=\kappa$. Then $S(\beta)= \mathcal{Z}(\beta, \kappa)(\frac{1}{2\beta-1}+\frac{1}{2\beta}\ln\frac{r_c^2}{4\pi\alpha'}-\frac{\ln \kappa}{1-1/2\beta}+\frac{1}{1-1/2\beta}\ln \epsilon)$. Acting with Tsytlin's prescription, (with $\kappa=1$), we get $\frac{\partial}{\partial \ln \epsilon}S(\beta)=\frac{1}{2\beta-1}\mathcal{Z}(\beta)$. We are puzzled by what it means that the entropy is finite for all values of $\beta$ and even what it means to have $\Lambda$ and $\eps$ depend on $\beta$ in such a way.}
\begin{equation}\label{eq:Z_1_entropy}
S_1(r_c)= \FTP S_{\mathrm{IR}}(\eps)  = \frac{r_c}{\sqrt{\pi\alpha'}},
\end{equation}
which is finite! One should immediately make the crucial observation that it is a genuinely stringy mechanism—the moduli integral—that makes the entropy finite: unlike in local QFT, the moduli integration allowed us to correlate the IR with UV, allowing the entropy to be renormalized so that all divergences are removed and a finite result remains. 



\medskip

\paragraph{Comments.}

Let us now try to better understand what the choice $\Lambda \eps = \kappa^2$ actually means. 

Let $z \in \mathbb{C}$ be a small neighborhood of a nodal point (pinching point) in a degenerating surface.  Define  the annulus around the nodal point
\begin{equation}\label{eq:annulus-collar}
A(\epsilon,\Lambda)\;=\;\bigl\{\,z\in\mathbb{C}\ \big|\ \epsilon \le |z| \le \Lambda\,\bigr\}\,,
\end{equation}
where $\epsilon>0$ and $\Lambda>0$ are the inner and outer radii, respectively. 

There are \emph{two distinct} choices to make: (1) a \emph{gauge (coordinate)} choice that fixes the overall scale of the local chart $z$, and (2) what amounts to a \emph{renormalization scheme} choice that relates the inner and outer radii.

There is an obvious scaling freedom: one may uniformly rescale the local coordinate by a constant factor
\begin{equation}\label{eq:annulus-dilation}
z\ \mapsto\ a\,z\qquad (a>0)\,,
\end{equation}
which takes $(\epsilon,\Lambda)\mapsto (a\,\epsilon,\ a\,\Lambda)$. The \emph{ratio}
\begin{equation}\label{eq:annulus-intrinsic-ratio}
u = 2\pi \log \frac{\Lambda}{\epsilon}\,,
\end{equation}
is left invariant under \eqref{eq:annulus-dilation}); $u$  measures how far apart the two boundary circles are irrespective od the the actual values of $\Lambda$ and $\eps$.

Because $z\mapsto a z$ with $a>0$ scales both radii together, we are free to fix the radius of \textit{any} circle as we rescale both ends of the annular region. A convenient choice is to pin the circle whose radius is the \emph{geometric mean} of the two boundaries
\begin{equation}\label{eq:step2-mid-radius}
r_\star\;\equiv\;\sqrt{\epsilon\,\Lambda}\,,
\end{equation}
to a fixed \emph{reference radius} $\kappa>0$
\begin{equation}\label{eq:step2-gauge}
r_\star\;=\;\kappa\,.
\end{equation}

\medskip

Independently, we can specify how the two ends of the annulus may be related. One choice is to hold \eqref{eq:step2-gauge} fixed while pushing the boundaries apart
\begin{equation}\label{eq:step2-scheme}
\epsilon\,\Lambda\;=\;\kappa^{2} \ \qquad\text{(scheme choice)}.
\end{equation}
\eqref{eq:step2-scheme} says that as the outer boundary recedes and the inner one shrinks, they do so in a way that treats the two sides of the annulus on equal footing. With the choice \eqref{eq:step2-gauge} in place, one can remove the divergences associated from the degeneration limit by taking the limit
\begin{equation}\label{eq:step2-limit}
\Lambda\to\infty,\qquad \epsilon\to 0,\qquad \epsilon\,\Lambda=\kappa^{2}\,,
\end{equation}
which keeps the reference circle $|z|=\kappa$ fixed. Since no physical observable depends on the value of $\kappa$. we choose the map $z\mapsto z/\kappa$, after which the reference circle is the unit circle $\epsilon\,\Lambda\;=\;1 \,.$

\medskip

Now we comment on the linear dependence of \eqref{eq:Z_1_entropy} on $r_c$. It is clear that the entropy vanishes right at the tip $r=0$ and blows up as $r_c \to \infty$. We are not sure if this is a bug or feature of our calculation but we will try to make sense of it here.  


So why does the entropy vanish when we take $r_c = 0$? Technically speaking, the solution of the equations of motion for a string starting at $r=0$ acquires infinite action (because of the $1/r^2$ potential blows) and, therefore, $e^{-I_{min}}$ and $\mathcal{Z}_1$ vanishes. See \eqref{eq:K_1_classical} and \eqref{eq:Zmod_u}. The only finite-action configuration with the boundary condition $r(0) = r(\ell_{\tau}) = 0$ is the trivial one $r(\tau) = 0$ (when $N_{\text{def}}=0$). In the other limit where $r_c \gg 0$, the string endpoint is effectively in the flat region far from the conical deficit, and hence does not see the horizon at the tip; we are basically in the zero-defect sector \eqref{eq:K0_def} where the entropy vanishes, although we do not see it vanish from \eqref{eq:Z_1_entropy}.

This limiting behavior of the entropy strongly suggests that our semiclassical description is only valid in a certain range of $r_c$. However, within our current calculation, we do not have a diagnostic that allows us to determine this allowed range of $r_c$. Thus, while the linear dependence $r_c$ may hint at an intrinsic stringy density of states localized near the tip, our computation  does not know how to precisely determine in which region this interpretation is trustworthy. So overall, to be nonzero, the entropy we calculated requires a finite radial cutoff; it also vanishes if $\ell_s \rightarrow 0$, which suggests that $\alpha'$ is a natural UV regulator in string theory.

As we will also explain in Section \ref{sec:constant_time_slice}, the image of cylindrical worldsheet in target space is a 2D annulus\footnote{The worldsheet of the string sweeps the surface of a 2D annulus between $r_c$ and $R$ in the $(r,\theta)$ plane with area
$$
A_{\text{annlus}}=\int_0^{\ell_\tau}\!d\tau\int_0^{2\pi}\!d\sigma\;r\,|\partial_\tau r|
=2\pi\beta\!\int_{r_c}^{R}\!r\,dr
=\pi\beta\,(R^2-r_c^2)
=\pi\beta\,r_c^2\sinh^2 u,
$$
where $R=r_c\cosh u$. The Nambu--Goto action is
$$
I_{\text{cl}}(u)=\frac{1}{2\pi\alpha'}A_{\text{annulus}}
=\frac{\beta\,r_c^2}{2\alpha'}\sinh^2 u.\
$$.} with area $\pi (R^2 - r^2_c)$, where $r_c$ is the radius of the inner disk cut around the tip at $r=0$ and $R$ is the radius of the outer disk. The open string lives on a half-line on a constant Euclidean $\theta$-slice of the cone. Sending $r_c \to 0$ collapses the 2D annulus to a punctured-disk or, equivalently, the line defect to a point, which immediately results in a vanishing entropy.

Inspired by the cigar, we may naively try to identify $\frac{r_c}{\ell_s} \sim \frac{1}{g_s^2} = e^{-2\Phi_0}$, where $\Phi_0$ is the constant mode of the dilaton. Then, we would have to conclude that the string coupling blows up right at the tip and goes to zero as $r_c \to \infty$ when the entropy is supposed to vanish (because the $1/r^2$ potential also goes to zero). This also implies that the region $r_c \approx 0$ can never be trusted since the semiclassical computation ceases to be make sense in this strong coupling regime.


\medskip

A reader familiar with Tseytlin's $\FTP$ off-shell prescription would probably observe that we have used it in a way that is markedly different from how it has traditionally been used in the literature, e.g. in \cite{Ahmadain:2022tew,Ahmadain:2022eso}. Normally, we would have acted on $\mathcal{Z}_1 (\beta, \Lambda, \eps)$ \eqref{eq:IR_integral_behavior_new}, to remove the logarithmic divergence and obtain a finite (renormalized) object. However, $\mathcal{Z}_1 (\beta, \Lambda, \eps)$ does not have any logarithmic divergence to begin with. Rather, we directly acted on $\eqref{eq:SIR_beta1_eps_only}$ \textit{after} going on-shell with $\beta =1$; it is only then that we had an idea what to do. We think that $\mathcal{Z}_1 (\beta, \Lambda, \eps)$, being a function of an \textit{arbitrary} value of $\beta$, is an object that can in principle be far off-shell, in RG space, from Rindler space at $\beta =1$. On the other side, Tseytlin's off-shell prescription is valid \textit{only} when the off-shell background is nearby a string background. So, we believe that it only makes sense to apply Tseytlin's prescription to \eqref{eq:SIR_beta1_eps_only}. 

\medskip

So far, we have focused on the case for one line defect. The case for $N_{\mathrm{def}}\in\mathbb{Z}_{\ge 1}$ defects is straightforward
\begin{equation}\label{eq:tail_many_beta}
e^{-I_{\min}^{(N_{\mathrm{def}})}(\ell_\tau)}
\sim C^{\,N_{\mathrm{def}}}\,\ell_\tau^{-\frac{N_{\mathrm{def}}\,a}{2}}\,.
\end{equation}

The amplitude $\mathcal{Z}_{N_{\text{def}}}(\beta,\Lambda, \eps)$ $\displaystyle \int^{\infty} d\ell_\tau\, = e^{-I_{\min}^{(N_{\mathrm{def}})}(\ell_\tau)}$ has the following divergence structure
\begin{equation}\label{eq:cases_betaN}
\begin{aligned}
&\beta=\frac{N_{\mathrm{def}}}{2} \;\Longrightarrow\; \text{logarithmic IR divergence},\\
&\beta>\frac{N_{\mathrm{def}}}{2} \;\Longrightarrow\; \text{power-law IR divergence }
\propto \Lambda^{\,1-\frac{N_{\mathrm{def}}}{2\,|\beta|}}\quad(\Lambda\to\infty), \\
&\beta<\frac{N_{\mathrm{def}}}{2} \;\Longrightarrow\; \text{converges}\,.
\end{aligned}
\end{equation}
So we find that as $N_{\mathrm{def}}$ increases, the convergence bound $\beta<N_{\mathrm{def}}/2$ relaxes, and thus it becomes easier for $\mathcal{Z}_{N_{\text{def}}}$ to converge for sufficiently many line defects. Consequently, the entropy is different from that one-defect case; it does not generalize. It can be easily checked that when $\beta=1$ and if we insist on the same regularization scheme \eqref{eq:RG_scheme_UV_IR}, the modulus integral in $\mathcal{Z}_{N_{\text{def}}}$ converges for $N_{\text{def}}\geq 3$. Moreover, $C(\beta,\epsilon)\rightarrow 0$ when $\epsilon\rightarrow0$, so the entropy vanishes. When $N_{\text{def}}=2$, $\mathcal{Z}_2\sim -\epsilon \ln \epsilon$, so, again the entropy vanishes.

\section{On the notion of a target space time–slice wavefunctional from the worldsheet}\label{sec:constant_time_slice}
Here, we show that the radial trajectory (open string) of a set of equally-spaced (along the $\sigma$ direction) $N_{\text{def}}$ defect lines localized at $\sigma_k$ is necessarily defined on a constant Euclidean time ($\theta$-slice). Then, we argue that this approach, where we first integrate out the Euclidean time (as we did in \eqref{eq:theta_integral}), is a consistent worldsheet approach to define a state on a target-space Cauchy slice, in contrast to the usual method in the replica trick (see e.g. Section two in \cite{Solodukhin_2011} and Chapter 2 in \cite{Rangamani:2016dms}) which is not well-defined on the worldsheet.

\medskip

The 2D worldsheet action for the $\theta$–variable is
\begin{equation}\label{eq:I-theta}
I_\theta = \frac{\beta^2}{4\pi\alpha'}\int\!\!d\tau
\int_{0}^{2 \pi}\!\!d\sigma\;
r(\tau,\sigma)^2
\Big[(\partial_\tau\theta)^2+(\partial_\sigma\theta)^2\Big],
\end{equation}
has an equation of motion 
\begin{equation}\label{eq:EOM-theta}
\partial_\tau\!\big(r^2\,\partial_\tau\theta\big)
+\partial_\sigma\!\big(r^2\,\partial_\sigma\theta\big)
=0\,.
\end{equation}

Consider only worldsheet configurations with no $\tau$–winding in $\theta$ (which is reasonable because our line defects extend between the two sides of the cylinder in a straight line), then
\begin{equation}\label{eq:no-tau-winding}
\partial_\tau\theta(\tau,\sigma)\equiv 0\,,
\end{equation}
which is consistent with \eqref{eq:EOM-theta} and minimizes 
\eqref{eq:I-theta} because a nonzero $\partial_\tau\theta$ contributes 
a positive cost $\propto r^2(\partial_\tau\theta)^2$. 
With \eqref{eq:no-tau-winding}, \eqref{eq:EOM-theta} reduces to
\begin{equation}\label{eq:EOM-theta-reduced}
\partial_\sigma\!\big(r(\tau,\sigma)^2\,\partial_\sigma\theta(\sigma)\big)=0\,.
\end{equation}
In a $\sigma$-interval around any defect, the radial trajectory satisfies $\partial_{\sigma} r = 0$. Then plugging in $r(\tau)$ into \eqref{eq:EOM-theta-reduced} gives
\begin{equation}\label{eq:theta-sigma-linear}
r(\tau)^2\,\partial_\sigma\theta(\sigma)
=\text{const}
\quad\Longrightarrow\quad
\partial_\sigma\theta(\sigma)=\frac{2\pi W}{2 \pi} = W\,,
\end{equation}
where $|W| = N_{\text{def}}\in\mathbb{Z}$.
Integrating \eqref{eq:theta-sigma-linear} gives
\begin{equation}\label{eq:theta-profile}
\theta(\sigma)
=\theta_0+ N_{\text{def}} \, \sigma\,.
\end{equation}
Evaluating \eqref{eq:theta-profile} on the defect worldline at $\sigma=\sigma_k=2\pi k/N_{\text{def}}$ 
yields (choosing $\sigma_0$ = 0 and $\theta_0 =0$)
\begin{equation}\label{eq:theta-on-line}
\theta(\tau,\sigma_k)
=\theta_k:= N_{\text{def}} \, \sigma_k \,=2\pi k\equiv 0\mod 2\pi.
\end{equation}
We then conclude that with \eqref{eq:theta-on-line}, the radial trajectory of a set of equally-spaced (along the $\sigma$ direction) $N_{\text{def}}$ defect lines localized at $\sigma_k$ is necessarily defined on a constant Euclidean time ($\theta$-slice)
\begin{equation}\label{eq:radial-worldline}
\big(r(\tau,\sigma_k),\theta(\tau,\sigma_k)\big)
=\big(r(\tau),\theta_k\big),\qquad 0\le\tau\le\ell_\tau.
\end{equation}
Observe that we never have to specify or fix which constant–$\theta$ slice $r(\tau)$ lives on because it does not matter for the construction: by integrating $\theta$ out, we effectively pick an \textit{arbitrary} constant-Euclidean time slice; this is similar to how we define a pure state (density matrix) on a Cauchy slice to compute entanglement entropy in a local QFT using the replica trick. 

\begin{figure}
\centering
\includegraphics[width=0.4\linewidth]{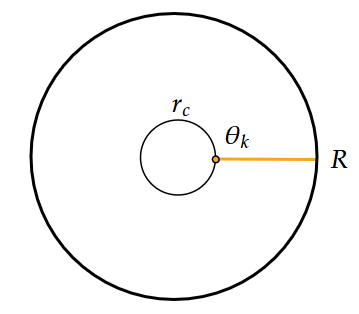}
\caption{The radial trajectory $r(\tau)$ of a set of equally-spaced $N_{\text{def}}$ defect lines (open strings) localized at $\sigma_k$ necessarily  lives on a constant-Euclidean-time slice of the cone at $\theta_k$. In our construction, this is implemented by integrating over $\theta$ everywhere except at the cylinder boundaries. This means we never have to fix the cone slice in advance: it can be \emph{any} slice. The figure shows one such slice.} 
\label{fig:annulus}
\end{figure}

\medskip

In a local QFT, the replica trick is the path integral approach to computing entanglent entropy (and more generally the Renyi entropies). The entangling state is defined by a Euclidean path integral with a specified boundary condition on a Cauchy slice $\Sigma|_{\theta = 0}$, a codimension-one surface in spacetime. Let $\phi$ be a scalar field with Euclidean action $I[\phi]$, and $x^\mu$ the spacetime coordinates. The path integral over the half-line region $\mathcal{A}$ with boundary data $\phi_0$ (a wavefunctional) defines the vacuum state
\begin{equation}\label{eq:QFT-state}
\Psi\!\left[\phi_0(x)\right]
\;=\; \int_{\phi|_{\theta = 0}=\phi_0}\!\!\!\mathcal{D}\phi\; e^{-I[\phi]}\,.
\end{equation}
We will argue below that \eqref{eq:QFT-state} on a string worldsheet raises a conceptual issue and argue that the method we used in this paper, where we first integrate out Euclidean time is a well-defined alternative on the worldsheet.

\medskip

In string theory, the dynamical variables are the embeddings $X^\mu:\Sigma\to\mathcal{M}$ of the worldsheet $\Sigma$ into target space $\mathcal{M}$. A cut in the target space (a Cauchy slice) is specified as the level set of a function $F:\mathcal{M}\to\mathbb{R}$, namely $F(X)=0$, which is then pulled back along $X^\mu$. Equivalently, the \emph{preimage} of the cut on the worldsheet is the locus (where $\sigma$ here denotes both $\sigma$ and $\tau$ on the worldsheet)
\begin{equation}\label{eq:preimage}
C \;:=\; \big\{\sigma\in\Sigma \;\big|\; F\big(X(\sigma)\big)=0\big\}\,.
\end{equation}
For example, in two target-space dimensions with polar coordinates $(\theta,r)$, one obvious choice is $F(X)=\theta$ so that $F=0$ defines the radial slice $\theta=0$.

At the path integral level (with no additional insertions), a putative wavefunctional of the cut would be a functional of $F$; na\"ively one should sum over all possible curves $C\subset\Sigma$
\begin{equation}\label{eq:sum_naive}
\Psi[F]\;\stackrel{\text{na\"ively}}{=}\;\sum_{C\subset\Sigma}\;
\int_{\;F(X)=0\ \text{on }C}\!\!\!\!\!\!\!\!\!\!\mathcal{D}X\; e^{-I[X]}\;=\;\int\mathcal{D}X\; e^{-I[X]}\,.
\end{equation}
The last equality holds because every target space configuration $X(\sigma)$ determines a unique $C$ of the level set; thus the sum over $C$ simply reconstructs the wavefucntional without the $F(x) = 0$ boundary condition. Consequently, the resulting $\Psi[F]$ is \emph{independent} of the location of the cut, which is \textit{not} the desired notion of a target space Euclidean-time slice wavefunctional. See \cite{Balasubramanian:2018} for an attempt to define state on the Cauchy slice in loop-space of open string field theory.

In this work, we propose an approach which is well-defined on the worldsheet. We integrate out the Euclidean thermal circle and, as a result, obtain a local one-dimensional line action $I_{\text{def}}[X;\gamma_k]$ for each $\gamma_k$ localized at $\sigma_k$ on the worldsheet; each $\gamma_k$ extends between the two boundary components of the cylinder with its two endpoints constrained to map to the Cauchy slice $F=0$ in target space. In other words, we instead \emph{fix} the preimage curve \eqref{eq:preimage} by inserting a prescribed set of codimension-one line defects. The insertion of the operator that encodes these defects implements the preimage $C=\cup_k\gamma_k$ by construction: the endpoints are frozen at $r_c$, and the induced (effective) $I_{\text{def}}$ provides an energetic cost for worldsheet configurations that deviate from that choice, thereby turning the would-be sum over $C$ into a functional that depends on the location of the Cauchy slice in the target space.

With the defects in place, the state on a Cauchy slice in 2D, is defined by the same worldsheet path integral but with the defect action included (with $X=r$), namely\footnote{Note that this is equivalent to imposing the Dirichlet boundary conditions on the fluctuation field $\eta(\ell_{\tau}, \sigma)$ and $\eta(0, \sigma)$ at $r=r_b$. See \eqref{eq:eta_BC_prod}.}
\begin{equation}
\Psi_{F}\,[\text{edge data}]:=\int\mathcal{D}r\;\exp\{-I_{\text{free}}[r]-\sum_k I_{\text{def}}[r;\gamma_k]\} \, \prod_\sigma \delta\left(r\left(\ell_\tau, \sigma\right)-r_c\right)
\end{equation}
Here \textit{edge data} denotes the boundary values of the embedding fields at the defect endpoints (i.e. the data on the preimage of the Cauchy slice, which is simply $r_c$ in the 2D case). Because $I_{\text{def}}$ emerges \textit{after} integrating out the Euclidean thermal circle on the slice, it depends on the geometric parameters that specify the cut (e.g., on $r_c$ and the cone angle $\beta$ in our work) and on the locations of the endpoints along the boundaries; consequently, $\Psi_{F}$ is sensitive to the choice of target space slice $F=0$. In practice, this construction avoids the degeneracy of the na\"ive sum over preimages in \eqref{eq:sum_naive} and yields a conceptually clear definition of the entangling state directly from the worldsheet: the defects serve as the preimage of the Cauchy slice, while the induced 1D dynamics (the line defect action) encodes the edge degrees of freedom that the local QFT construction accounts for by boundary conditions in spacetime.

\section{The Semiclassical Entropy from the NLSM of a 2D Cone }\label{sec:SC_saddle_Circular}

In this section, we revisit the same NLSM on the flat 2D cone, but follow a different route: we do \emph{not} discretize the worldsheet and we do \emph{not} use a Hubbard–Stratonovich transformation. Instead, we reduce the full two–dimensional worldsheet action to a one–dimensional one by using an ansatz  $r(\tau,\sigma)=r(\tau)$ and $\theta(\tau,\sigma)=W\sigma$ with $W\in\mathbb{Z}$, subject to Dirichlet boundary conditions $r(0)=r(\ell_\tau)=r_c>0$. We compute the classical (off-shell) action and use it to compute the semiclassical partition function in the long–cylinder IR regime. We then  calculate the entropy by varying $\beta$ and find it to be finite in \textit{each} winding sector $W$ (which we use in this section instead of $N_{\text{def}}$) and has a maximum at $r_c = \sqrt{\alpha'}/|W|$. We find that the $|W|=1$ sector is dominant and the maximum entropy evaluates to a non-universal constant $\sim 2\log 2$. When we sum over all winding modes $|W|$, the entropy has a maximum finite value $4 \log 2$ even when $r_c\rightarrow 0^+$ and converges in the UV limit but only for $r_c > 0$.

\subsection{A 1D action with Dirichlet boundary conditions}
We work with the original 2D cone NLSM
\begin{equation}\label{eq:P_action_def}
I_{\text{ws}}[r,\theta]
=\frac{1}{4\pi\alpha'}\int_{0}^{\ell_\tau}\!d\tau\int_{0}^{2\pi}\!d\sigma\;
\Big[(\partial_\tau r)^2+(\partial_\sigma r)^2+\beta^{2}r^{2}\big((\partial_\tau\theta)^2+(\partial_\sigma\theta)^2\big)\Big],
\end{equation}
with Dirichlet boundary conditions $r(0,\sigma)=r(\ell_\tau,\sigma)=r_c>0$ and $\theta\sim\theta+2\pi$. 
We use the ansatz 
\begin{equation}\label{eq:circ_gauge}
r(\tau,\sigma)=r(\tau),\qquad 
\theta(\tau,\sigma)=W\sigma,\qquad W\in\mathbb{Z},
\end{equation}

For any smooth embedding whose $\theta$-map is locally bijective, worldsheet diffeomorphisms allow us to choose coordinates so that $\sigma$ parametrizes the target angle monotonically. The residual reparametrization freedom on the $\sigma$–circle can then be used to straighten the map $S^1_\sigma\to S^1_\theta$ to a linear representative, yielding $\theta(\tau,\sigma)=W\,\sigma$ with $W\in\mathbb{Z}$ fixed by the periodicities of $\theta$ and $\sigma$. With this choice, the action in each winding sector is minimized when $r$ is only $\tau$-dependent, so the stationary configuration satisfies $r(\tau,\sigma)=r(\tau)$. The assumption here is that $\theta(\cdot,\sigma)$ is \textit{locally} one–to–one in $\sigma$; non-bijective maps of $\theta$ (e.g.\ $\theta(\sigma)=\sigma^2$) are thus outside this class.

The ansatz \eqref{eq:circ_gauge} turns a two–dimensional variational problem into a one–dimensional ordinary differential equation boundary–value problem for $r(\tau)$. With \eqref{eq:circ_gauge}, the target space image of a $\sigma$–circle is a round circle (so no ripples) of circumference $2\pi\,\beta\,|W|\,r(\tau)$; it is the level set at fixed $\tau$. Also, the $\theta$–map is single–valued: $\theta(\tau,\sigma+2\pi)-\theta(\tau,\sigma)=2\pi W$.


Plugging \eqref{eq:circ_gauge} into \eqref{eq:P_action_def} and integrating over $\sigma$ gives the 1D (boundary) action
\begin{equation}\label{eq:P_reduced_action}
I_{\text{1D}}[r]=\frac{1}{4\pi\alpha'}\int_{0}^{\ell_\tau}\!d\tau\int_{0}^{2\pi}\!d\sigma\;\big[\dot r^{\,2}+\beta^2 r^2 W^2\big]
=\frac{1}{2\alpha'}\int_{0}^{\ell_\tau}\!d\tau\;\big[\dot r(\tau)^{2}+\omega^2\,r(\tau)^2\big],
\end{equation}
where we defined
\begin{equation}\label{eq:omega_def}
\omega\;:=\;\beta\,|W|\ \ \ge 0,\qquad \dot r:=\frac{dr}{d\tau}.
\end{equation}
The Euler--Lagrange equation is
\begin{equation}\label{eq:EOM_circ}
\frac{d}{d\tau}\Big(\frac{1}{\alpha'}\,\dot r\Big)-\frac{1}{\alpha'}\,\omega^2 r=0
\qquad\Longrightarrow\qquad
\ddot r(\tau)-\omega^2\,r(\tau)=0\,.
\end{equation}
Because $L$ has no explicit $\tau$--dependence, the (Euclidean) energy is given by
\begin{equation}\label{eq:energy_const}
\mathcal{E}:=\dot r\,\frac{\partial L}{\partial \dot r}-L
=\frac{1}{2\alpha'}\big(\dot r^{\,2}-\omega^2 r^2\big)
\quad\text{is constant.}
\end{equation}
The minimum halfway along the radial trajectory is
\begin{equation}\label{eq:rmin_def}
\rho:=r\!\left(\frac{\ell_\tau}{2}\right),\qquad \dot r\!\left(\frac{\ell_\tau}{2}\right)=0,\qquad 0<\rho<r_c,
\end{equation}
(the potential $\propto r^2$ drives $r$ inward). Evaluating \eqref{eq:energy_const} at $\tau=\ell_\tau/2$ gives
\begin{equation}\label{eq:energy_value}
\mathcal{E}=-\frac{1}{2\alpha'}\,\omega^2\rho^2.
\end{equation}
Equating \eqref{eq:energy_const} and \eqref{eq:energy_value} yields the \emph{first integral}
\begin{equation}\label{eq:first_integral}
\dot r(\tau)^2=\omega^2\big(r(\tau)^2-\rho^2\big)\,.
\end{equation}
On the interval $0\le\tau\le \ell_\tau/2$, $r$ rises from $\rho$ to $r_c$, so we take $\dot r\ge 0$ and write
\begin{equation}\label{eq:separate_vars}
\frac{dr}{\sqrt{r^2-\rho^2}}=\omega\,d\tau.
\end{equation}
Integrating \eqref{eq:separate_vars} from $\tau=\ell_\tau/2$ (where $r=\rho$) to $\tau$ yields
\begin{equation}\label{eq:sol_integral}
\int_{\rho}^{r(\tau)}\frac{dr}{\sqrt{r^2-\rho^2}}=\omega\!\int_{\ell_\tau/2}^{\tau}d\tau'
\quad\Longrightarrow\quad
\operatorname{arcosh}\!\Big(\frac{r(\tau)}{\rho}\Big)=\omega\Big(\tau-\frac{\ell_\tau}{2}\Big).
\end{equation}
Exponentiating, we obtain 
\begin{equation}\label{eq:r_profile}
r(\tau)=\rho\,\cosh\!\Big(\omega\,(\tau-\tfrac{\ell_\tau}{2})\Big)\,,\qquad 0\le\tau\le\ell_\tau,
\end{equation}
and by symmetry the same also holds on $(\ell_\tau/2,\ell_\tau]$. 
Imposing the Dirichlet boundary conditions $r(0)=r(\ell_\tau)=r_c$ fixes $\rho$
\begin{equation}\label{eq:rho_fix}
r_c=\rho\,\cosh\!\Big(\frac{\omega\ell_\tau}{2}\Big)
\quad\Longrightarrow\quad
\rho=\frac{r_c}{\cosh(\frac{\omega\ell_\tau}{2})}\,.
\end{equation}
As a sanity check, the slope is
\begin{equation}\label{eq:slope}
\dot r(\tau)=\rho\,\omega\,\sinh\!\Big(\omega\,(\tau-\tfrac{\ell_\tau}{2})\Big)
\quad\Rightarrow\quad
\frac{\dot r}{r}=\omega\,\tanh\!\Big(\omega\,(\tau-\tfrac{\ell_\tau}{2})\Big),
\end{equation}
which is what we expect.

\medskip

For any solution of \eqref{eq:EOM_circ},
\begin{equation}\label{eq:total_deriv_identity}
\frac{d}{d\tau}\big[r(\tau)\,\dot r(\tau)\big]=\dot r^{\,2}+r\,\ddot r=\dot r^{\,2}+\omega^2 r^2.
\end{equation}
Hence
\begin{equation}\label{eq:Imin_boundary}
I\big[r_{\text{cl}}\big]
=\frac{1}{2\alpha'}\int_{0}^{\ell_\tau}\!\big(\dot r^{\,2}+\omega^2 r^2\big)\,d\tau
=\frac{1}{2\alpha'}\,\Big[r(\tau)\,\dot r(\tau)\Big]_{\tau=0}^{\tau=\ell_\tau}.
\end{equation}
Using \eqref{eq:r_profile} and \eqref{eq:rho_fix},
\begin{align}
r(0)&=r(\ell_\tau)=r_c,\qquad
\dot r(0)=-\,\rho\,\omega\,\sinh\!\Big(\tfrac{\omega\ell_\tau}{2}\Big),\qquad
\dot r(\ell_\tau)=+\,\rho\,\omega\,\sinh\!\Big(\tfrac{\omega\ell_\tau}{2}\Big),\label{eq:bdry_vals}\\
\end{align}
Then
\begin{align}
I\big[r_{\text{cl}}]
\frac{1}{2\alpha'}\cdot 2 r_c \,\rho\,\omega\,\sinh\!\Big(\tfrac{\omega\ell_\tau}{2}\Big)
=\frac{r_c^2\,\omega}{\alpha'}\,\tanh\!\Big(\tfrac{\beta |W| \, \ell_\tau}{2}\Big),\label{eq:Imin_final}
\end{align}
where we used $\rho=r_c/\cosh(\frac{\omega\ell_\tau}{2})$ and \eqref{eq:omega_def}.

Thus, the classical off-shell stationary-point action is\footnote{We could have also derived $I_{\text{min}}$ by direct substitution. Inserting \eqref{eq:r_profile} into $I_{\text{1D}}$ and using
$\dot r^{\,2}+\omega^2 r^2=\rho^2\omega^2\big(\sinh^2+\cosh^2\big)=\rho^2\omega^2\cosh(2\omega(\tau-\ell_\tau/2))$. Then Integrating and using \eqref{eq:rho_fix} gives the same minimal action.} 
\begin{equation}\label{eq:Imin_boxed}
I_{\min}(\ell_\tau\beta, W, r_c)
=\frac{\beta\,|W|\,r_c^2}{\alpha'}\;\tanh\!\Big(\frac{\beta\,|W|\,\ell_\tau}{2}\Big).
\end{equation}
which is strictly positive for $(\ell_\tau>0,\ \omega>0)$.\footnote{It is important to note that $I_{\text{min}}$ is \textit{not} an on-shell saddle since it does \textit{not} solve the Virasoro constraints $\left(\partial_\tau r\right)^2-\omega^2 r^2=\omega^2\left(r^2-\rho^2\right)-\omega^2 r^2=-\omega^2 \rho^2<0$. The constraint is not satisfied except in the trivial limits $\rho=0$ i.e. $r \rightarrow 0$, or $\omega=0$ i.e. $W=0$. One way to satisfy the Virasoro constraints with $W\neq 0$ and with Dirichlet $r(0,\sigma)=r(\ell_\tau,\sigma)=r_c>0$, the \textit{only} pointwise solutions are the exponentials $r(\tau)=A e^{ \pm \omega \tau}$; these cannot reach the same positive value at both ends, (as $\partial_\tau r$ will be discontinuous, for example a cusp). In other words, there is no smooth embedding with $W \neq 0$ and $\partial_\tau \theta=0$. However, with these boundary conditions, we would be computing a 2-point scattering amplitude, not a partition function. This is because the unfixed end of the string will not have any place to terminate at infinity, which implies the string propagates in only one direction and is not able to turn around.}

\medskip

\begin{figure}
\centering
\includegraphics[width=0.6\linewidth]{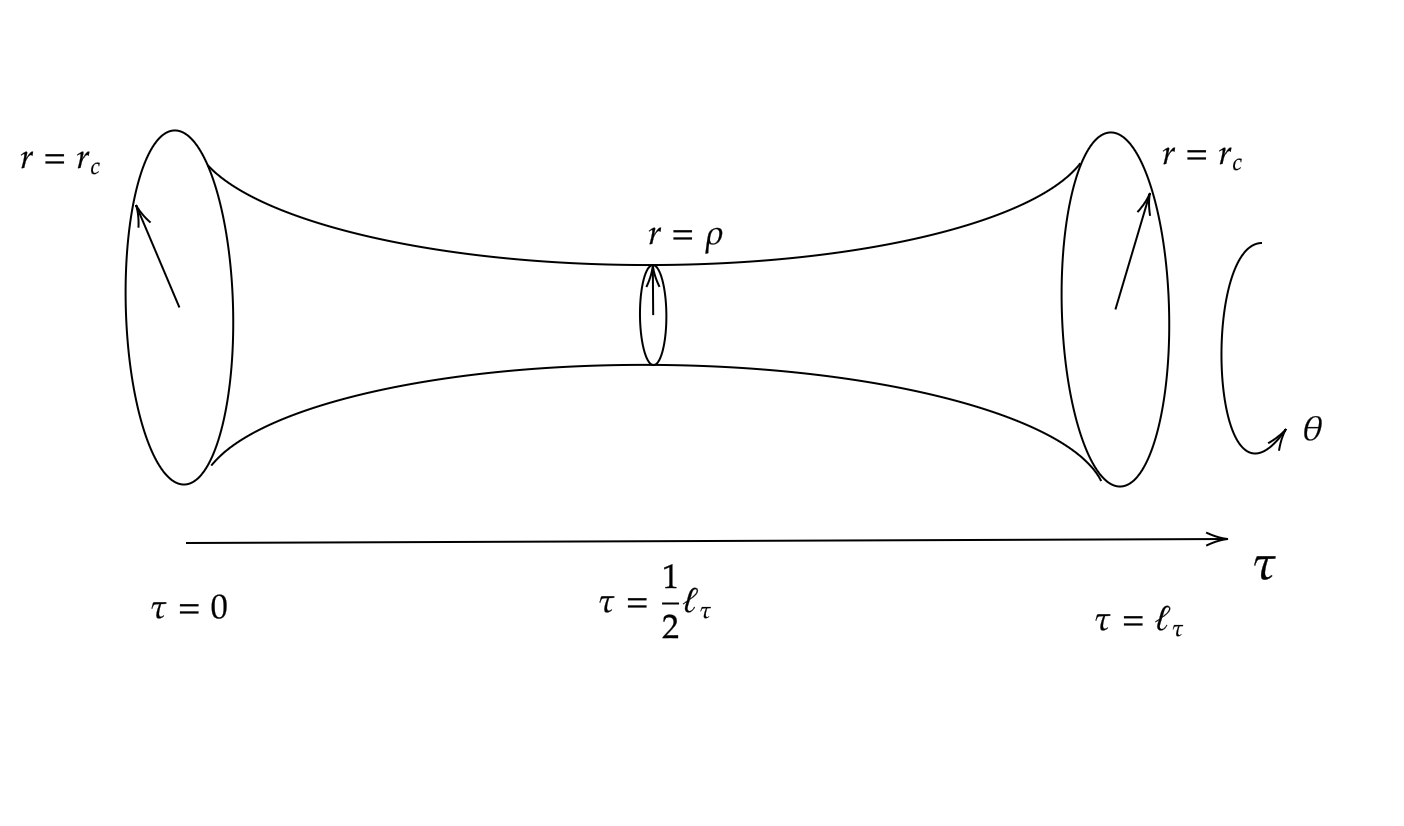}
    \caption{A 3D projection of the of the embedding map. The target space image of the worldsheet is a a double trumpet with two ends of radius $r_c$ at $\tau=0$ and $\tau=\ell_\tau$ joined by a narrow throat of radius $\rho$ at $\tau=\ell_\tau/2$. The slope and curvature make the shape manifest: $\dot r(\tau)=\rho\,\omega\,\sinh\!\big(\omega(\tau-\ell_\tau/2)\big)$ vanishes at the throat and grows monotonically away from it, while $\ddot r(\tau)=\omega^2 r(\tau)>0$ everywhere, so the shape is strictly convex and smoothly decreases from $r_c$ to $\rho$ and then increases back to $r_c$. The width of the throat is controlled by the dimensionless parameter $\omega\ell_\tau$: in the long–cylinder limit $\ell_\tau\to\infty$, one has $\rho=\tfrac{r_c}{\cosh(\omega\ell_\tau/2)}\sim 2\,r_c\,e^{-\omega\ell_\tau/2}$, producing an exponentially thin neck, whereas for short cylinders $\rho\to r_c$ and the trumpet turns into a cylinder. For $|W|>1$, it is $W$-sheeted surface along the $\theta$-direction.}
    \label{fig:trumpet}
\end{figure}

It is clear from \eqref{eq:rho_fix} that the image of each $\tau$-slice is a round circle of circumference $2\pi\,\beta\,|W|\,r(\tau)$. In three dimensions, the embedding map $X:(\tau, \sigma) \mapsto(r(\tau), \theta(\sigma))$ ($W=1$) is a double trumpet with two mouths of radius $r_c$ at $\tau=0$ and $\tau=\ell_\tau$ joined by a narrow throat of radius $\rho$ at $\tau=\ell_\tau/2$. See fig. \eqref{fig:trumpet}. In two dimensions, the map is a $W$-sheeted annulus $\left\{(r, \theta): 0 <\rho \leq r \leq r_c, \hspace{0.4em} \theta \sim \theta+2 \pi\right\}$ along $\theta$. Observe that, in contrast to the annulus in Section~\ref{sec:constant_time_slice} (from fig. \eqref{fig:annulus}) which chops off the area near the tip of the cone, the annulus excises the cone outside a small neighborhood of the tip, so the string never explores the region $r>r_c$.

\medskip

\subsection{The amplitude and entropy}
\label{subsec:PF_entropy}
 
Define the positive quantities
\begin{equation}
\label{eq:defs_ABC}
A\;\coloneqq\;\frac{|W|\,r_c^2}{\alpha'}\,,
\qquad
b\;\coloneqq\;\frac{|W|\,\ell_\tau}{2}\,.
\end{equation}

The partition function $K(\beta, W,\ell_\tau)$ in each winding sector $W$ is
\begin{equation}
\label{eq:f_def}
K(\beta, W,\ell_\tau)\;=\;\exp\!\Big[-\beta\,A\,\tanh(\beta b)\Big].
\end{equation}

\medskip

We normalize the $\widehat Z(\beta,W,r_c)$ so that it vanishes at $\beta =1$
\begin{equation}
\label{eq:Zhat_def}
\widehat Z(\beta,W,r_c)
=\int_{0}^{\infty}\!d\ell_\tau\;\Big[K(\beta,W,\ell_\tau)-\beta K(1,W,\ell_\tau)\Big].
\end{equation}

The entropy is
\begin{equation}
\label{eq:S_replica_def}
S(W,r_c)\;=\;\bigl(1-\beta\,\partial_\beta\bigr)\,\widehat Z(\beta,W,r_c)\,\Big|_{\beta=1}\,.
\end{equation}
Since $\widehat Z(1,W,r_c)=0$, only the derivative contributes
\begin{equation}
\label{eq:S_as_df}
S(W,r_c)\;=\;\;\int_{0}^{\infty}\!d\ell_\tau\;\Big[(1-\beta\partial_\beta) K(\beta,W,\ell_\tau)\Big]_{\beta=1}.
\end{equation}
From \eqref{eq:f_def}, we have
\begin{equation}
\label{eq:df_dbeta}
\partial_\beta K(\beta,W,\ell_\tau)
=K(\beta,W,\ell_\tau)\;\partial_\beta\!\Big[-\beta A\,\tanh(\beta\, b)\Big]
=-\,K(\beta,W,\ell_\tau)\,A\Big[\tanh(\beta\, b)+\beta b\,\sech^2(\beta\, b)\Big].
\end{equation}
Evaluating at $\beta=1$ (so $K(1,W,\ell_\tau)=e^{-A\tanh b}$) yields 
\begin{equation}
\label{eq:S_exact_integrand}
\quad
S(W,r_c)\;=\;\int_{0}^{\infty}\!d\ell_\tau\;e^{-A\,\tanh b}\,\Big[1\;+\;A\,\tanh b\;+\;A\,b\,\sech^2 b\Big].
\quad
\end{equation}
We split $S(W,r_c)$ into divergent and finite terms
\begin{equation}
\label{eq:S_split}
S(W,r_c)=S_{\text{div}}(W,r_c)+S_{\text{finite}}(W,r_c),
\end{equation}
where
\begin{align}\label{eq:S_finite}
S_{\text{div}}:=\int_{0}^{\infty}\!d\ell_\tau\;e^{-A\,\tanh b}(1+\,A\,\tanh b\,), \quad 
S_{\text{finite}}:=\int_{0}^{\infty}\!d\ell_\tau\;e^{-A\,\tanh b}\,A\,b\,\sech^2 b.
\end{align}
The divergent piece is the result of the entropy operator directly acting on the linear part of the action, i.e., on $(\beta \, A)$ but not $\tanh \, \beta b$ in \eqref{eq:f_def}. In what follows, we drop $S_{\text{div}}$ and keep the finite piece $S_{\text{finite}}$, but we will come back later to study the effect of $S_{\text{div}}$ on the behavior of the entropy especially in the UV and IR limits.

\medskip

Using $d\ell_\tau=\frac{2}{|W|}\,db$ and the change of variables $u=\tanh b$ ($du=\sech^2 b\,db$, $b=\operatorname{arctanh}u$), we obtain 
\begin{equation}
\label{eq:S_edge_exact}
S_{\text{finite}}(W,r_c)
=\frac{2A}{|W|}\int_{0}^{\infty}\!db\;b\,\sech^2 b\,e^{-A\tanh b}
=\frac{2A}{|W|}\int_{0}^{1}\!du\;\operatorname{arctanh}u\;e^{-A u},
\end{equation}
which is manifestly finite for all $A>0$.

On the support of $\sech^2 b$, we may replace $e^{-A\tanh b}\to e^{-A}$ (errors are exponentially small for large $b$). Using the elementary identity
\begin{equation}
\label{eq:int_ln2}
\int_{0}^{\infty}\!db\;b\,\sech^2 b
=\Big[b\tanh b-\log\cosh b\Big]_{0}^{\infty}=\log 2,
\end{equation}
we find using \eqref{eq:defs_ABC}
\begin{equation}
\label{eq:S_edge_IR}
S_{\text{finite}}(W,r_c)\;\simeq\;\frac{2A}{|W|}\,e^{-A}\,\log 2
=2\log 2\;\frac{r_c^2}{\alpha'}\,e^{-\;|W|\,r_c^2/\alpha'}\;,
\end{equation}
So, the entropy is finite in each winding sector $|W|$. For fixed $W\neq 0$, \eqref{eq:S_edge_IR} has a maximum
\begin{equation}
\label{eq:sector_max}
\frac{d}{d(r_c^2)}\Big[(r_c^2)\,e^{-|W|r_c^2/\alpha'}\Big]=0
\quad\Longrightarrow\quad
r_c^2=\frac{\alpha'}{|W|},\qquad
S_{\max}(W)=\frac{2\log 2}{e}\,\frac{1}{|W|}.
\end{equation}
We see the \textit{dominant }contribution comes from the  $|W|=1$ sector and $S_{\max}(1) \sim 2\log 2$! The entropy peaks at the string scale $r_c=\sqrt{\alpha'}$.

\medskip

We can sum over \textit{all} nontrivial winding sectors. Summing \eqref{eq:S_edge_IR} over nonzero windings (and accounting for the two possible orientations of each $|W|$) gives
\begin{equation}
\label{eq:S_total_BE}
S(r_c)
=2\sum_{W\in\mathbb{Z}\setminus\{0\}}S_{\text{finite}}(W,r_c)
=4\log 2\,\frac{r_c^2}{\alpha'}\sum_{W=1}^{\infty}e^{-W r_c^2/\alpha'}
=4\log 2\;\frac{r_c^2}{\alpha'}\;\frac{1}{e^{\,r_c^2/\alpha'}-1}\;.
\end{equation}

Let us revisit the divergent piece $S_{\mathrm{div}}$ in \eqref{eq:S_finite} and analyze in what it changes \eqref{eq:S_total_BE}. We first analyze the IR regime of the integral, where $\ell_\tau \gg 1$. In this limit, $\tanh b \simeq 1$, and each winding mode reduces to a constant contribution
\begin{equation}
e^{-A}(1+A) \;=\; e^{-|W| r_c^2/\alpha'}\bigl(1+|W| r_c^2/\alpha'\bigr)\,.
\end{equation}
The corresponding IR divergence is therefore linear in $\Lambda$, the upper limit of the integral. If we sum over all winding modes, the constant part of the integrand becomes
\begin{equation}
2\,\frac{e^{r_c^2/\alpha'}\bigl(1+r_c^2/\alpha'\bigr)-1}{\bigl(e^{r_c^2/\alpha'}-1\bigr)^2}\,,
\end{equation}
which is finite for $r_c \neq 0$. Thus, even after including all winding contributions, the linear divergence in $\Lambda$ remains unchanged. \footnote{We think this divergence arises from the bulk tachyon in bosonic string theory and can be removed by analytic continuation or a local counterterm.}

The UV behavior can be analyzed in a similar way. For a fixed $W$, the integrand remains finite for all $\ell_\tau \in [0,\infty)$, and hence, there is no UV divergence. When we sum over $W$, the behavior depends on $r_c$. Using $\tanh b \approx b$, the integrand becomes
\begin{equation}
\left(1 - k \frac{\partial}{\partial k}\right)\vartheta_3(0,k)\bigg|_{k = \frac{r_c^2 \ell_\tau}{2\alpha'}},
\end{equation}
where
\begin{equation}
  \vartheta_3(z,q) \;=\; \sum_{n=-\infty}^{\infty} q^{n^2} e^{2 i n z},
  \qquad |q| < 1 \,.
\end{equation}
When $r_c = 0$, the integral is rapidly divergent. For $r_c > 0$ and $\ell_\tau \ll 1$, we have $k \ll 1$, and $\vartheta_3$ diverges as $\ell_\tau \to 0^+$. This can be understood as follows: $\ell_\tau = 0$ corresponds to a degenerate limit in which the worldsheet collapses to a one-dimensional circle and the theory reduces to an ordinary QFT. The winding modes are then analogous to Fourier modes on a circle, and summing over all winding numbers reproduces the familiar UV divergence of the QFT.

However, in string theory, we must integrate over the full moduli space; the degenerate QFT limit has measure zero. The UV contribution of this corner is finite provided the integrand decays faster than $O(\ell_\tau^{-1})$ as $\ell_\tau \to 0^+$. This condition is indeed satisfied here by a modular transformation
\begin{equation}
\vartheta_3\bigl(0,e^{-k}\bigr) \;\approx\; \sqrt{\frac{\pi}{k}}
\quad (k \to 0^+),
\end{equation}
and we find that the integral $S_{\mathrm{div}}$ converges in the UV regime for $r_c \neq 0$. The overall effect is the addition of an $r_c$-dependent \textit{constant} to \eqref{eq:S_total_BE}, but we believe a more careful analysis of this contribution is required.

\medskip

We would like to make some remarks on the behavior of the entropy \eqref{eq:S_total_BE}. 

\medskip

Let $\varepsilon:=r_c^2/\alpha'>0$. Then, the sum over winding numbers in \eqref{eq:S_total_BE} takes the form of the average occupation number of a single bosonic mode in a single harmonic oscillator with respect to the Bose–Einstein distribution with energy $\varepsilon$ and inverse temperature $\beta =1$ (which normally is 2$\pi$)
\begin{equation}
\label{eq:nbar}
\bar n(\varepsilon)\;=\;\frac{1}{e^{\beta \,\varepsilon}-1}\,.
\end{equation}
With \eqref{eq:nbar}, the total entropy \eqref{eq:S_total_BE} is the \textit{mean} energy to wind once around the $r_c$-circle
\begin{equation}
\label{eq:S_as_energy_of_mode}
S(r_c)\;=\;4\ln 2\;\frac{\varepsilon}{e^{\varepsilon}-1}
\;=\;4\ln 2\;\epsilon\,\bar n(\varepsilon)\,,
\end{equation}

\medskip

\noindent Notice that $S(r_c)$ is strictly \textit{decreasing} with $r_c$
\begin{equation}
\label{eq:monotonicity}
\frac{d}{d \varepsilon}\!\left[\frac{\varepsilon}{e^\varepsilon-1}\right]
=\frac{e^\varepsilon-1-xe^\varepsilon}{(e^\varepsilon-1)^2}
=\frac{e^\varepsilon(1-\varepsilon)-1}{(e^\varepsilon-1)^2}\;<\;0\qquad(\varepsilon>0),
\end{equation}
and has the limits
\begin{equation}
\label{eq:limits}
S(r_c)\;=\;4\ln 2\;\Big[1-\tfrac{1}{2}\,\varepsilon+\tfrac{1}{12}\,\varepsilon^2+O(\varepsilon^3)\Big]
\quad(\varepsilon\to 0^+),\qquad
S(r_c)\;=\;4\ln 2\;\varepsilon\,e^{-\varepsilon}\,\Big[1+O(e^{-\varepsilon})\Big]
\quad(\varepsilon\to\infty).
\end{equation}
Thus, we see $S(r_c)$ is finite and approaches a maximum $4\ln 2$ as $r_c\to 0^+$ (in contrast to the entropy \eqref{eq:Z_1_entropy} which vanishes as $r_c\to 0$), while for $r_c\gg\sqrt{\alpha'}$ it is exponentially suppressed: $S(r_c)\;=\;4\ln 2\;\varepsilon\,e^{-\varepsilon}\,\Big[1+O(e^{-\varepsilon})\Big]\xrightarrow{\varepsilon\to\infty}0$. In the small–$r_c$ regime (where $\varepsilon=\frac{r_c^c}{\alpha^{\prime}} \ll 1$), many winding sectors contribute but the series ultimately converges to \eqref{eq:S_total_BE}; For $r_c\gtrsim\sqrt{\alpha'}$, the $|W| \gg 1$ sectors are exponentially suppressed, and the total entropy is dominated by that the $|W|=1$ sector: $S(r_c) \sim 4 \log 2$.

\section{Discussion and Future Directions}\label{sec:discussion}
We would like to make a few comments here on several aspects of our work in this paper. We do this in no particular order.





\medskip


As mentioned in the introduction, because the worldsheet topology is fixed, our setup only permits vortex insertions at the two boundaries (two punctures) where the winding number can change and the string pinches off at the conical tip, as a result. If the topology was not fixed, there would be additional \textit{bulk} worldsheet configurations that should be accounted for: flux lines that begin in the interior and either terminate in the interior or at one of the two boundaries. In target space, they would be strings that cross the codimension-2 entangling surface an infinite number of times without a proper regulator; then we would be computing the amplitude of an $n$-punctured sphere. The stiffness term suggested in \cite{ahmadain2022off} is one example of how to regulate the number of string crossings. Relatedly, taking into those bulk configurations would potentially allow us to probe the BKT transition on the worldsheet. We leave this future work.



The recent work~\cite{Balasubramanian:2025zey,Balasubramanian:2025hns} (see also \cite{Colafranceschi:2023moh} for a potentially related work) has shown that the Euclidean gravitational path integral with a periodic Euclidean time circle does \emph{not} compute the thermal trace in perturbative gravity because the Hilbert space does not factorize on its own. In order to factorize, one must include nonperturbative effects: wormholes connecting the asymptotic boundaries, equivalently, the sum over topologies. In string theory, the Hagedorn temperature  \cite{AtickWitten:1988} in flat spacetime, or Berezinskii--Kosterlitz--Thouless (BKT) transition \cite{Sathiapalan:1986,Kogan:1987} in 2D string theory, where there is a proliferation of vortex–antivortex pairs (holes) on the worldsheet, is a genuinely non-perturbative phenomenon. It has long been conjectured by Susskind \cite{Susskind-Speculations:1993,Susskind-Lorentz:1993} that the thermal ensemble of strings at very high temperatures would be dominated, for entropic reasons, by long strings. According to this picture near the horizon, at radius $\sim \ell_s$ there exists a single long string.  See \cite{Mertens:Thesis:2015} for a comprehensive discussion. Could the two non-perturbative behaviors be related in any sensible way?

It is also important to understand, at a deeper level, the renormalization scheme we used to eliminate the IR divergence in the entropy. In our computation, the combination $(\sqrt{\Lambda}\sqrt{\epsilon})$ plays a crucial role in producing a finite answer, but at present, it is unclear whether this structure is an accident of our 2D toy model or a sign of something more universal. It would be important to determine how general this cancelation mechanism is: does the same scheme extend to higher–dimensional target spaces, to supersymmetric strings, or to higher–genus worldsheets where moduli spaces are more complicated? Understanding this could shed more light on the universality of tip–localized degrees of freedom and their contribution to black hole entropy.



\medskip

\medskip

The line configurations we found in this work may be interpreted as open strings emitted and reabsorbed by a localized defect at $r_c$. Roughly speaking, the defect can be interpreted as an \textit{off-shell D0-brane} \cite{Rychkov:2002ni} in which the string is temporarily excited into the bulk but constrained to return to the defect point. The defect at $r_c$ represents a fixed background which plays the role of a source or sink for worldsheet excitations, analogous to a D0-brane in critical string theory but defined within a non-conformal, off-shell setting. In this sense, these open string configurations, we argue, are \textit{boundary-localized} degrees of freedom associated with the interaction of the worldsheet with the point defects at $r_c$. Perhaps these degrees of freedom are best interpreted as \emph{edge modes} rather than conventional open strings. It would be very interesting to study this connection further.

While we have computed the entropy in this work in an off-shell setting for general values of $\beta$, there is a way to compute the black hole on-shell at $\beta=1$. Here is a rough idea of how to do that.

The string matter partition function
\begin{equation}
\ln Z(\beta)
=
\int \mathcal{D}r\,[\beta r]\;\mathcal{D}\theta\;
\exp\!\left[-\int \mathrm{d}^2\sigma\,\Big(\partial_a r\,\partial^a r + \beta^{2} r^{2}\,\partial_a \theta\,\partial^a \theta\Big)\right],
\end{equation}
has a measure factor with a Jacobian $\sqrt{\det G}=[\beta r]$. The entropy is then computed, as usual, by varying the cone angle $\beta$
\begin{equation}
S=\Big(1-\beta\,\partial_\beta\Big)\,\ln Z(\beta)\,\Big|_{\beta=1}.
\end{equation}
We can get rid of the $\beta$–dependence of the measure by rescaling $r$
\be
r'=\sqrt{\beta}\,r.
\ee
In terms of $r'$, the target space metric reads
\be
\mathrm{d}s^{2}
=\frac{1}{\beta}\,\mathrm{d}r'{}^{2}+\beta\, r'{}^{2}\,\mathrm{d}\theta^{2},
\ee
whose determinant is \emph{independent} of $\beta$. Therefore, the $\beta$–variation of the measure drops out, and the entropy reduces to the variation of the action alone
\begin{equation}
S=\left\langle \int \mathrm{d}^{2}\sigma\,\Big(-\,\partial r\,\partial r + r^{2}\,\partial \theta\,\partial \theta \Big)\right\rangle.
\end{equation}

The author in \cite{Lin:2017uzr} discusses how the RT area term may potentially arise from additional gravitational degrees of freedom localized at the entangling surface in an emergent gauge theory. Although our system is not a gauge theory (and we do not yet see how a gauge theory description would emerge), it is nevertheless intriguing to view our construction through the lens of the emergent gauge theory/edge-mode analogy.

There are several directions which we have not explored in this paper that we would like to study in future work. In Part II, we will present a detailed computation of $K_1$ \eqref{eq:K_1_alpha'} where we integrate over the string fluctuation $\eta(\sigma, \tau)$. We will implement Dirichlet boundary conditions by keeping only the $\sin$ modes in the expansion of the string coordinates rather than by Lagrange multipliers. Note that in this case the fluctuations around the minimum (the open string with no crossings), will inevitably cross below $r=r_c$ a number of times such that that the number of left crossings minus the number right crossings differs by 1. Nothing forbids these intermediate returns; they are simply weighted by the action. Although the path integral weight makes steps toward smaller $r$ costly, the string will occasionally dip below $r_c$ - those are the finite number of horizon crossings along the way. In a configuration with $N_{\text{def}}$ insertions, we have $N_{\text{def}}$ such independent open strings, each with endpoints pinned at $r_c$.

It would be interesting to see the effect of $\alpha'$-corrections and compute the 1-loop correction (torus amplitude) to the entropy, which is expected to diverge in bosonic string theory because of the tachyon and vanish for type-II strings. We would also like to include extra target space dimensions $\mathbb{R}^{D-2}\times\text{(cone)}$, and to apply our construction to $\text{AdS}_3$ NLSM and attempt a derivation of the Ryu-Takayanagi holographic entanglement entropy formula \cite{RTPhysRevLett.96.181602, HRT:2007} in semiclassical gravity a la Lewkowycz-Maldacena \cite{LM2013,FLM2013}\footnote{The recent work in \cite{Wu:2025qwc} appears to be related to ours and a step in this direction.}. The authors of \cite{Brustein:2021qkj} solved a reduced version of the 2D Horowitz–Polchinski equations of motion. It would be interesting to understand whether and in what sense, their solution is related to our 1D half-line solution. 

Ultimately, our hope is that the work in this paper can be used as a step towards a stringy realization of the ER= EPR conjecture \cite{Maldacena-Cool:2013,Jafferis-ER=EPR:2021}. This requires an in-depth understanding of string theory in Lorentzian signature and time-dependent backgrounds (e.g. bag-of-gold spacetime), and how or if it is possible to build a smooth spacetime from disconnected components.

\acknowledgments
We thank Shoaib Akhtar for early collaboration on this project. We thank Aron Wall, Hong Liu, Alexander Frenkel, Yiming Chen, Ronak Soni, Wen-Xin Lai, Jun Liu, Prem Kumar, Sean Hartnoll for insightful discussions. We are especially grateful for Aron Wall and Hong Liu for their valuable comments and feedback, and Alex Frenkel for critical review of the draft.  AA is supported by STFC Consolidated Grant No. ST/X000648/1. MY is supported by a Gates Scholarship (\#OPP1144).

\noindent\footnotesize \textbf{Open Access Statement} - For the purpose of open access, the authors have applied a Creative Commons Attribution (CC BY) licence to any Author Accepted Manuscript version arising. \\
\textbf{Data access statement:} no new data were generated for this work

\bibliographystyle{JHEP}
\bibliography{main.bib}

\appendix

\section{\texorpdfstring{The $K_0$ partition function}{The K0 partition function}}\label{app:K_0}

In this appendix we present the full mathematical details for calculating $K_0$, including the fluctuations at the quadratic level.

The cylinder partition function $K_0$ for $N_{\text{def}} = 0$ (i.e.\ zero defect lines) is
\begin{align} 
 K_0(\ell_\tau; r_b, r_b)
 &= \int \left[\prod_{m,n}
      \frac{\sqrt{\alpha'}\,\mathrm{d}a_{m,n}}{4\pi \beta r_0}\right]
    \left[\prod_{n}\mathrm{d}\Lambda_n\right]
    \exp\!\left[i r_b \Lambda_0\right]
    \nonumber\\
 &\qquad \times
    \exp\!\left[
      -\sum_{m,n}
        \Big(a_{m,n}+ i\frac{\Lambda_{-n}}{M_{m,n}A_\Sigma}\Big)
        \Big(a_{-m,-n}+ i\frac{\Lambda_{n}}{M_{m,n}A_\Sigma}\Big)
        M_{m,n}A_\Sigma
      -\frac{\Lambda_n\Lambda_{-n}}{A_\Sigma M_{m,n}}
    \right]
    \nonumber\\[0.5em]
 &= \prod_{m,n}
    \frac{\sqrt{\alpha'/A_\Sigma}}{4\beta r_0\, M_{m,n}}
    \int\left[\prod_n \mathrm{d}\Lambda_n\right]\exp\!\left[
      -\frac{r_b^2}{4\sum_{m\neq 0}\frac{1}{A_\Sigma M_{m,0}}}
    \right]\nonumber \\
    &\qquad \times\exp\!\left[
      -\sum_n
        \Big(\sum_m \tfrac{1}{M_{m,n}A_\Sigma}\Big)
        \Big(\Lambda_n - \tfrac{i r_b \delta_{0,n}}
                          {2\sum_{m\neq 0}\frac{1}{A_\Sigma M_{m,n}}}\Big)
        \Big(\Lambda_{-n} - \tfrac{i r_b \delta_{0,n}}
                          {2\sum_{m\neq 0}\frac{1}{A_\Sigma M_{m,n}}}\Big)
    \right]
    \nonumber\\
 &= \prod_{m,n}
    \frac{\sqrt{\alpha'/A_\Sigma}}{4\beta r_0\, M_{m,n}}\;
    \prod_n
    \frac{\pi}{
      \displaystyle\sum_m \frac{1}{A_\Sigma M_{m,n}}
    }\;
    \exp\!\left[
      -\frac{r_b^2}{4\sum_{m\neq 0}\frac{1}{A_\Sigma M_{m,0}}}
    \right]\!,
 \label{eq:K0-def}
\end{align}
where $M_{m,n}$ are the eigenvalues of the worldsheet kinetic operator evaluated on the cylinder and $A_\Sigma$ is the area of the spatial slice.

Next, we will process each term in \eqref{eq:K0-def}.

\subsection{First term}

The first term can be isolated as the (regularized) divergent piece
\begin{equation}\label{eq:2.46}
 \left(4\beta r_0\sqrt{\frac{A_\Sigma}{\alpha'}}\right)^{-NM+1}
 \prod_{(m,n)\neq(0,0)} M_{m,n}^{-1/2}
 \;=\;
 \beta\, F,
\end{equation}
where $M$ and $N$ are, respectively, the numbers of spatial sites and Matsubara modes, and $F$ (after regularization) is indeed $\beta$-independent
\be
F
 = \left(4 r_0\sqrt{\frac{A_\Sigma}{\alpha'}}\right)
   \prod_{(m,n)\neq(0,0)} M_{m,n}^{-1/2}.
\ee
The `$-1$' in the exponent of \eqref{eq:2.46} excludes the overall zero mode.

\medskip

We can perform the product over $m$ first using that,
\begin{equation}
    \prod_m M_{m,n}
    = 4\sinh^2\!\left(
        \frac{\ell_\tau}{2}\sqrt{n^2-\mu^2}
      \right),
\end{equation}
where $\mu^2 = 2\pi\alpha'/r_0^2$.

In evaluating products of constants, we use the symmetric zeta regularization
$1 + 2\zeta(0) = 0$. Concretely, for any constant $c$ (independent of $m,n$)
\begin{equation}
    \prod c
    = c^{\sum_{m\in \mathbb{Z}} 1}
    = c^{\,2\zeta(0)+1}
    = 1.
    \label{eq:sym-reg}
\end{equation}
The same symmetric scheme can be applied to the overall prefactor in the first term of
\eqref{eq:2.46}, so it becomes simply
$4\beta r_0\sqrt{A_\Sigma/\alpha'}$. This reproduces the familiar Susskind--Uglum statement that the zero-winding sector scales as
$K_0 \propto \beta$, and therefore \emph{does not contribute to the entropy.}

The absolute value of the full double product
$\prod_{m,n\in\mathbb Z} M_{m,n}$ depends on the regularization scheme. Its
\emph{$\mu$-dependence}, however, is canonical. We package this dependence in the ratio
\begin{equation}\label{eq:D-def}
\mathcal{D}(\mu)
:=\frac{\displaystyle\prod_{m,n\in\mathbb Z}
       \Big(n^2+\big(\tfrac{2\pi m}{\ell_\tau}\big)^2-\mu^2\Big)}
        {\displaystyle\prod_{m,n\in\mathbb Z}
       \Big(n^2+\big(\tfrac{2\pi m}{\ell_\tau}\big)^2\Big)}.
\end{equation}
Any remaining overall normalization of $\prod M_{m,n}$ is then a $\mu$-independent constant
$\mathcal{N}$:
\[
\prod_{m,n} M_{m,n} \;=\; \mathcal{N}\,\mathcal{D}(\mu).
\]
One may regard \eqref{eq:D-def} as a closed-form definition without further regularization. However, this can be rewritten involving the more familiar Jacobi theta functions by regarding $\mathcal D(\mu)$ as an even entire function of $\mu$ with $\mathcal D(0)=1$, and determine it
from its zero set. The zeros of $\mathcal D$ occur precisely when one of the numerator factors vanishes
\begin{equation}\label{eq:zeros-mu}
\quad
\mu=\pm\,\omega_{m,n},
\qquad
\omega_{m,n}:=\sqrt{\,n^2+\Big(\tfrac{2\pi m}{\ell_\tau}\Big)^2\,},
\qquad (m,n)\in\mathbb Z^2.
\quad
\end{equation}
Each such zero is simple in $\mu$ (for $\omega_{m,n}\neq0$) because
$\partial_\mu\big[n^2+(\tfrac{2\pi m}{\ell_\tau})^2-\mu^2\big]\big|_{\mu=\pm\omega_{m,n}}=-2\mu\neq 0$.
\footnote{Degeneracies can occur when distinct lattice points $(m,n)$ yield the same $\omega_{m,n}$, in which case the multiplicity is the number of such representations.}

Because $\mathcal D$ is even, entire, and normalized by $\mathcal D(0)=1$, its Weierstrass factorization in $\mu$
can be written as
\begin{equation}\label{eq:weierstrass-mu}
\mathcal D(\mu)
=\prod_{(m,n)\in\mathbb Z^2}
\left(1-\frac{\mu^2}{\omega_{m,n}^2}\right)\,e^{\;\mu^2/\omega_{m,n}^2},
\end{equation}
where the (entire) exponential factors tame convergence and encode a $\mu$-independent normalization.\footnote{In the
symmetric/zeta scheme used throughout, these exponential factors resum to a $\mu$-independent constant, so they can be
dropped in the ratio $\mathcal D(\mu)$. Equivalently, one can work with the canonical Hadamard factors for an even
order-2 entire function and absorb the resulting constant in the overall regularization.}

Let us now introduce the torus variables
\begin{equation}\label{eq:A9}
z:=\frac{\ell_\tau\mu}{2},\qquad \tau:=i\,\frac{\ell_\tau}{2\pi}\quad(\Im\tau>0).
\end{equation}
In these variables, the zero set \eqref{eq:zeros-mu} is the image, under $\mu\mapsto z=\ell_\tau\mu/2$, of the real axis
points $z=\pm\frac{\ell_\tau}{2}\sqrt{\,n^2+(\tfrac{2\pi m}{\ell_\tau})^2\,}$. On the other hand, the odd Jacobi theta function
$\vartheta_1(z|\tau)$ is the \emph{canonical} entire function whose simple zeros form the complex lattice
$\{\,z=\pi(n+m\tau)\,:\,n,m\in\mathbb Z\,\}$. Matching these two descriptions is achieved by the identity
\begin{equation}
\vartheta_1(z|\tau)
=2q^{1/4}\sin z\prod_{k=1}^\infty(1-q^{2k})\bigl(1-2q^{2k}\cos 2z+q^{4k}\bigr),
\qquad q:=e^{i\pi\tau}=e^{-\ell_\tau/2}.
\end{equation}
Since $\vartheta_1(z|\tau)\sim\vartheta_1'(0|\tau)\,z$
as $z\to0$, normalizing by $\mathcal D(0)=1$ gives
\begin{equation}\label{eq:D-theta-again}
\mathcal D(\mu)
=\left[\frac{\vartheta_{1}\!\Big(\tfrac{\ell_\tau\mu}{2}\,\Big|\, i\,\tfrac{\ell_\tau}{2\pi}\Big)}
              {(\tfrac{\ell_\tau\mu}{2})\,\vartheta_{1}'\!\Big(0\,\Big|\, i\,\tfrac{\ell_\tau}{2\pi}\Big)}\right]^2.
\end{equation}

Putting all of of the above together, the first term becomes
\begin{equation}
    4\beta r_0\sqrt{\frac{A_\Sigma}{\alpha'}}\,
    \frac{
      \big(\tfrac{\ell_\tau\mu}{2}\big)\,
      \vartheta_{1}'\!\Big(0 \,\Big|\, i\,\tfrac{\ell_\tau}{2\pi}\Big)
    }{
      \vartheta_{1}\!\Big(\tfrac{\ell_\tau\mu}{2}\,\Big|\, i\,\tfrac{\ell_\tau}{2\pi}\Big)
    },
\end{equation}
with $\mu^2 = 2\pi\alpha'/r_0^2$.

\subsection{Second term}

Using the Mittag--Leffler expansion \cite{ML}, for $n\neq 0$ the exponential factor in
\eqref{eq:K0-def} involves
\begin{equation}\label{eq:2.56}
    S_n := \sum_m \frac{1}{A_\Sigma M_{m,n}}
    \;=\;
    \frac{1}{A_\Sigma}
    \frac{2\pi \ell_\tau}{\kappa_n}\,
    \coth\!\left(\frac{\kappa_n \ell_\tau}{2}\right),
\end{equation}
where
\[
\kappa_n^2 := n^2 - 2\pi \alpha'/r_0^2 = n^2 - \mu^2.
\]
For $n=0$ we must omit the $m=0$ mode
\begin{equation}\label{eq:2.57}
    S_0 := \sum_{m\neq 0} \frac{1}{A_\Sigma M_{m,0}}
    \;=\;
    \frac{1}{A_\Sigma}
    \left[
    -\frac{4\pi}{\mu^2}
    \;+\;
    \frac{2\pi\,\ell_\tau}{\mu}\,
    \cot\!\Big(\frac{\ell_\tau\mu}{2}\Big)
    \right].
\end{equation}
The second term in \eqref{eq:2.57} is the (zeta-regularized) product $\prod_{n=-\infty}^{\infty} S_n$. We again
use \eqref{eq:sym-reg} to drop $\mu$-independent constants such as
$A_\Sigma$, $\ell_\tau$ and $\pi$. Define
\begin{equation}
\mathcal{S}(\mu)
:= \prod_{n\ge 0}^{\zeta} S_n(\mu),
\qquad
\text{and normalise by}\qquad
\frac{\mathcal{S}(\mu)}{\mathcal{S}(0)}.
\end{equation}

For fixed $n$, the $m$-spectrum on a circle of length $\ell_\tau$ gives
\begin{equation}
\prod_{m\in\mathbb Z}
\Big[\Big(\tfrac{2\pi m}{\ell_\tau}\Big)^2+\kappa_n^2\Big]
=4\sinh^2\!\Big(\frac{\ell_\tau \kappa_n}{2}\Big),
\end{equation}
and hence
\begin{equation}
\sum_{m\in\mathbb Z}
\frac{1}{\big(\tfrac{2\pi m}{\ell_\tau}\big)^2+\kappa_n^2}
=
\frac{\ell_\tau}{2\kappa_n}\,
\coth\!\Big(\frac{\ell_\tau\kappa_n}{2}\Big).
\end{equation}
Therefore
\begin{equation}
S_n(\mu)
= 4\pi\,
\partial_{\kappa_n^2}
\log\sinh^2\!\Big(\frac{\ell_\tau \kappa_n}{2}\Big).
\end{equation}

We use the torus variables \eqref{eq:A9}, as before. From the Jacobi product formulae and $\vartheta_{1}'(0|\tau)=2\pi\,\eta(\tau)^3$, the
$\mu$-dependence of the double product
$\prod_{m,n}\big(n^2+(2\pi m/\ell_\tau)^2-\mu^2\big)$ can be written as
\begin{equation}
\mathcal{D}(\mu)
=
\left[
\frac{\vartheta_{1}\!\big(z\,\big|\,\tau\big)}
     {z\,\vartheta_{1}'\!\big(0\,\big|\,\tau\big)}
\right]^{2}.
\end{equation}
Differentiating $\log\mathcal{D}$ with respect to $\mu^2$ reproduces
$\sum_{n} S_n$ (up to the known $n=0$ adjustment), and integrating back fixes the
$\mu$-dependence of $\prod S_n$ modulo an overall $\mu$-independent constant.

Carrying out the symmetric zeta regularization for the scale factors and using the identities above, one finds
\begin{equation}
\frac{\displaystyle\prod_{n\ge 0}^{\zeta} S_n(\mu)}
     {\displaystyle\prod_{n\ge 0}^{\zeta} S_n(0)}
=
\frac{\sin z}{z}\,
\left[
\frac{\vartheta_{1}\!\big(z\,\big|\,\tau\big)}
     {2\pi\,\eta(\tau)^{3}\,z}
\right]^{-1}
=
\frac{\sin\!\big(\tfrac{\ell_\tau \mu}{2}\big)}{\tfrac{\ell_\tau \mu}{2}}\,
\left[
\frac{\vartheta_{1}\!\Big(\tfrac{\ell_\tau \mu}{2}\,\Big|\, i\,\tfrac{\ell_\tau}{2\pi}\Big)}
     {2\pi\,\eta\!\Big(i\,\tfrac{\ell_\tau}{2\pi}\Big)^{3}\,
      \tfrac{\ell_\tau \mu}{2}}
\right]^{-1}.
\end{equation}

\subsection{Combining the first two terms}

Since many factors cancel, it is convenient to combine the first two terms at once and, using \eqref{eq:sym-reg}), drop all $\mu$-independent constants.

The first product may be written as
\begin{equation}
 \prod_{n\neq 0}
 \frac{1}{
   2\sinh\!\left(
     \frac{\ell_\tau}{2}\sqrt{n^2-\mu^2}
   \right)
 }
 \times
 \frac{
   \sqrt{\mu^2/4\pi}
 }{
   2\sin\!\left(\frac{\ell_\tau \mu}{2}\right)
 }.
\end{equation}

The inverse square of the second product is
\begin{equation}
 \prod_{n\neq 0}
 \frac{\coth(\kappa_n \ell_\tau/2)}{\kappa_n}
 \times
 \left[
 \frac{2}{\mu^2\ell_\tau}
 -
 \frac{1}{\mu}\,
 \cot\!\Big(\frac{\ell_\tau\mu}{2}\Big)
 \right].
\end{equation}

Combining them
\begin{equation}
 \prod_{n\neq 0}
 \left[
   \frac{\kappa_n}{\sinh(\kappa_n \ell_\tau)}
 \right]^{1/2}
 \times
 \big[\text{$n{=}0$ piece}\big]
 =
 \frac{1}{\sqrt{2}}\,
 \left[
 \frac{\sin\!\bigl(\ell_\tau \mu\bigr)}{\mu}\;
 \frac{\eta\!\left(i\,\tfrac{\pi}{\ell_\tau}\right)^{3}}
 {\vartheta_{1}\!\left(\ell_\tau \mu\,\middle|\,
                        i\,\tfrac{\pi}{\ell_\tau}\right)}
 \right]^{\!1/2}
 \times
 \big[\text{$n{=}0$ piece}\big].
\end{equation}
One convenient way to evaluate $\prod_n \kappa_n$ is via the Weierstrass product
\[
\frac{\sin(\pi \mu)}{\pi \mu}
= \prod_{n=1}^{\infty} \left(1-\frac{\mu^2}{n^2}\right),
\]
together with the zeta-regularisation $\prod_{n=1}^{\infty}n^2 = 2\pi$. Equivalently, dealing
with both terms together and using
\begin{equation}
 \prod_n
 \frac{\sinh(\ell_\tau \kappa_n)}
      {\ell_\tau \kappa_n}
 =
 \frac{
   \vartheta_1\!\big(i\ell_\tau a \,\big|\, e^{-\ell_\tau}\big)
 }{
   i\,\vartheta_1'\!\big(0\,\big|\, e^{-\ell_\tau}\big)
 },
\end{equation}
gives the same $\mu$-dependence. (Here $a$ is the relevant continuation parameter.)

\subsection{Final expression}

The last term in \eqref{eq:K0-def} is
\begin{equation}
 \exp\!\left[
   -\frac{r_b^2}{4\sum_{m\neq 0}\frac{1}{A_\Sigma M_{m,0}}}
 \right]
 =
 \exp\!\left[
   -\frac{A_\Sigma r_b^2}{
     4\left(
       -\frac{4\pi}{\mu^2}
       +
       \frac{2\pi\,\ell_\tau}{\mu}\,
       \cot\!\big(\tfrac{\ell_\tau\mu}{2}\big)
     \right)
   }
 \right]\!.
\end{equation}

Putting everything together, we get 
\be
\begin{aligned}
K_0(\beta,\ell_\tau, r_b)
 &=
 \left(4\beta r_0\sqrt{\frac{A_\Sigma}{\alpha'}}\right)
 \exp\!\left[
   \frac{A_\Sigma r_b^2}{
     4\left(
       \frac{4\pi}{\mu^2}
       -
       \frac{2\pi\,\ell_\tau}{\mu}\,
       \cot\!\big(\tfrac{\ell_\tau\mu}{2}\big)
     \right)
   }
 \right]
 \\
 &\quad \times
 \frac{1}{\sqrt{2}}\,
 \left[
 \frac{\sin\!\bigl(\ell_\tau \mu\bigr)}{\mu}\;
 \frac{\eta\!\left(i\,\tfrac{\pi}{\ell_\tau}\right)^{3}}
 {\vartheta_{1}\!\left(\ell_\tau \mu\,\middle|\,
                        i\,\tfrac{\pi}{\ell_\tau}\right)}
 \right]^{\!1/2}
 \times
 \big[\text{$n{=}0$ piece}\big].
\end{aligned}
\ee
Note that $A_\Sigma = 2\pi \ell_\tau$ enters explicitly, but the discretisation scale
$\epsilon$ does not. For the moduli integral and computing entropy from $K_0$, see subsection \ref{ssec:Cylinder_Ampl}.

\end{document}